\documentclass[12pt,twoside, a4paper]{article}
\def\pd{\partial}
\def\mc{\mathcal}

\usepackage[dvips]{graphicx}
\usepackage{amssymb}
\usepackage{amssymb,amsmath}
\usepackage{graphicx}
\usepackage{dsfont}
\usepackage{caption}
\usepackage{subcaption}
\input{epsf.sty}  
\begin{document}
\begin{center}
\LARGE{\textbf{Supersymmetric domain walls in $7D$ maximal gauged supergravity}}
\end{center}
\vspace{1 cm}
\begin{center}
\large{\textbf{Parinya Karndumri}$^a$ and \textbf{Patharadanai Nuchino}$^b$}
\end{center}
\begin{center}
String Theory and Supergravity Group, Department
of Physics, Faculty of Science, Chulalongkorn University, 254 Phayathai Road, Pathumwan, Bangkok 10330, Thailand
\end{center}
E-mail: $^a$parinya.ka@hotmail.com \\
E-mail: $^b$danai.nuchino@hotmail.com \vspace{1 cm}\\
\begin{abstract}
We give a large class of supersymmetric domain walls in maximal seven-dimensional gauged supergravity with various types of gauge groups. Gaugings are described by components of the embedding tensor transforming in representations $\mathbf{15}$ and $\overline{\mathbf{40}}$ of the global symmetry $SL(5)$. The embedding tensor in $\mathbf{15}$ representation leads to $CSO(p,q,5-p-q)$ gauge groups while gaugings in $\overline{\mathbf{40}}$ representation describes $CSO(p,q,4-p-q)$ gauge groups. These gaugings admit half-supersymmetric domain walls as vacuum solutions. On the other hand, gaugings involving both $\mathbf{15}$ and $\overline{\mathbf{40}}$ components lead to $\frac{1}{4}$-supersymmetric domain walls. In this case, the gauge groups under consideration are $SO(2,1)\ltimes \mathbf{R}^4$ and $CSO(2,0,2)\sim SO(2)\ltimes \mathbf{R}^4$. All of the domain wall solutions are analytically obtained. For $SO(5)$ gauge group, the gauged supergravity admits an $N=4$ supersymmetric $AdS_7$ vacuum dual to $N=(2,0)$ SCFT in six dimensions. The corresponding domain walls can be interpreted as holographic RG flows from the $N=(2,0)$ SCFT to non-conformal $N=(2,0)$ field theories in the IR. The solutions can be uplifted to eleven dimensions by using a truncation ansatz on $S^4$. Furthermore, the gauged supergravity with $CSO(4,0,1)\sim SO(4)\ltimes \mathbf{R}^4$ gauge group can be embedded in type IIA theory via a truncation on $S^3$. The uplifted domain walls, describing NS5-branes of type IIA theory, are also given.    
\end{abstract}
\newpage
%%%%%%%%%%%%%%%%%%%%%%%%%%%%%%%%%%%%%%%%%%%%%%%%%%%%%%%%%%%%%%%%%%%%%%%%%%%%%%%%%%%%%%%%%%%%%%%%%%%%%%%%%%%%%%%%%%%%%%%%%%%%%%%%%%%%%%%%%
\section{Introduction}
Supersymmetric $p$-branes have played an important role throughout the development of string/M-theory. These extended objects can be effectively described by using $(p+2)$-dimensional gauged supergravity (possibly including massive deformations in higher dimensions) in which they become domain walls. The latter are of particular interest in the DW/QFT correspondence \cite{DW_QFT1,DW_QFT2,DW_QFT3}, a generalization of the AdS/CFT correspondence \cite{maldacena}, and in cosmology, see for example \cite{DW_cosmology1,DW_cosmology2,DW_cosmology3}. In addition, classifications of supersymmetric domain walls can give some insight to the underlying structure of M-theory \cite{counting_DW} through the algebraic structure $E_{11}$ \cite{E11_West}. 
\\
\indent In ten dimensions, there is only one massive type IIA supergravity and hence only one possible domain wall \cite{Eric_DW_10D}. In nine and eight dimensions, half-supersymmetric domain walls have been studied in \cite{9D_DW_Cowdall,9D_DW} and \cite{8D_DW1,8D_DW2} using maximal gauged supergravities. In this paper, we will consider supersymmetric domain walls within maximal gauged supergravity in seven dimensions. General discussions about this type of solutions and examples of domain walls in $N=4$ gauged supergravity with $SO(5)$ gauge group have already been given in previous works \cite{Pope_DW_massive,DW_algebraic_curve,Eric_DW_maximal_SUGRA,Eric_SUSY_DW,Pope_symmetric_poten}. However, as pointed out in \cite{Eric_SUSY_DW}, a systematic study of these domain walls and explicit solutions in other gauge groups have not appeared so far. Similar solutions in lower-dimensional gauged supergravities can also be found in \cite{DW_Hull,4D_DW,3DN4_DW,3DN10_DW,2D_DW1,2D_DW2}. 
\\
\indent We will give a large number of supersymmetric domain wall solutions in maximal $N=4$ gauged supergravity with various gauge groups. The first $N=4$ gauged supergravity with $SO(5)$ gauge group has been constructed for a long time in \cite{7DN4_gauged1,7DN4_gauged2}. It can be obtained from a consistent truncation of eleven-dimensional supergravity on a four-sphere $S^4$ \cite{11D_to_7D_Townsend,11D_to_7D_Nastase1,11D_to_7D_Nastase2}. The most general deformations of the $N=4$ supergravity are obtained by using the embedding tensor formalism. These gaugings have been constructed in \cite{N4_7D_Henning}. There are two components of the embedding tensor transforming in $\mathbf{15}$ and $\overline{\mathbf{40}}$ representations of the global $SL(5)$ symmetry. As shown in \cite{Eric_SUSY_DW}, each of these components leads to half-supersymmetric domain walls. In addition, the $\mathbf{15}$ and $\overline{\mathbf{40}}$ parts give rise to domain walls supporting respectively tensor and vector multiplets on their world-volumes. Unlike higher-dimensional analogues, when both representations of the embedding tensor are present simultaneously, the domain walls are only $\frac{1}{4}$-supersymmetric. In this paper, we will analytically give solutions for domain walls of all these types. 
\\
\indent For gaugings in $\mathbf{15}$ representation, we will consider $CSO(p,q,5-p-q)\sim SO(p,q)\ltimes \mathbf{R}^{(p+q)(5-p-q)}$ gauge groups. For $SO(5)$ gauge group with known eleven-dimensional origin, solutions to $N=4$ gauged supergravity can be embedded in M-theory. Furthermore, this gauged supergravity also admits a maximally supersymmetric $AdS_7$ vacuum which is, according to the AdS/CFT correspondence, dual to $N=(2,0)$ superconformal field theory (SCFT) in six dimensions. The domain walls with an $AdS_7$ asymptotic can be interpreted as holographic RG flows from the $N=(2,0)$ SCFT to non-conformal field theories in the IR. We consider this type of domain walls in the context of the AdS/CFT correspondence and carry out their uplift to eleven dimensions. In addition, the gauging from $\mathbf{15}$ representation with gauge group $CSO(4,0,1)$ can be obtained from a truncation of type IIA supergravity on $S^3$ \cite{S3_S4_typeIIA}. We also give uplifted solutions of the domain walls from this gauge group in type IIA theory. 
\\
\indent For gaugings in $\overline{\mathbf{40}}$ representation, the gauge groups under consideration are $CSO(p,q,4-p-q)\subset SL(4)\subset SL(5)$. The existence of a higher-dimensional origin of the $SO(4)$ gauge group from a truncation of type IIB theory on $S^3$ has been pointed out in \cite{DW_QFT1}, and, recently, the corresponding truncation ansatz has been constructed in the framework of exceptional field theories in \cite{Henning_Emanuel}. Finally, for gaugings with the embedding tensor from both $\mathbf{15}$ and $\overline{\mathbf{40}}$ representations, we consider non-semisimple $SO(2,1)\ltimes \mathbf{R}^4$ and $SO(2)\ltimes \mathbf{R}^4$ gauge groups which give rise to $\frac{1}{4}$-supersymmetric domain walls.
\\
\indent 
The paper is organized as follow. In section \ref{N4_7D_SUGRA}, we give a review of the
maximal gauged supergravity in seven dimensions using the embedding tensor formalism. Half-supersymmetric domain walls for gauge groups $CSO(p,q,5-p-q)$ are given in section \ref{Y_gauging}. For $SO(5)$ gauge group, admitting a supersymmetric $AdS_7$ vacuum, we consider holographic RG flows from $N=(2,0)$ six-dimensional SCFT to non-conformal field theories in the IR and study an uplift to eleven dimensions of these solutions. Uplifted solutions to type IIA theory of domain walls in $CSO(4,0,1)$ gauge group are also given. We then perform a similar analysis for $CSO(p,q,4-p-q)$, $SO(2,1)\ltimes \mathbf{R}^4$ and $SO(2)\ltimes \mathbf{R}^4$ gauge groups in sections \ref{Z_gauging} and \ref{YZ_gauging}. Conclusions and comments on the results are given in section \ref{conclusion}. Consistent reduction ansatze for M-theory on $S^4$  and type IIA theory on $S^3$ which are useful to the discussion in the main text are reviewed in the appendix.

%%%%%%%%%%%%%%%%%%%%%%%%%%%%%%%%%%%%%%%%%%%%%%%%%%%%%%%%%%%%%%%%%%%%%%%%%%%%%%%%%%%%%%%%%%%%%%%%%%%%%%%%%%%%%%%%%%%%%%%%%%%%%%%%%%%%%%%%%
\section{Maximal gauged supergravity in seven dimensions}\label{N4_7D_SUGRA}
In this section, we give a brief review of $N=4$ gauged
supergravity in seven dimensions in the embedding tensor formalism. This section closely follows the original construction given in \cite{N4_7D_Henning} to which the reader is referred for more detail. 
\\
\indent The maximal $N=4$ supersymmetry consists of only the supergravity multiplet with the field content given by
\begin{equation}
(e^{\hat{\mu}}_\mu, \psi^a_\mu,A^{MN}_\mu,B_{M\mu\nu},\chi^{abc},{\mc{V}_M}^A) \, .
\end{equation}
Curved and flat space-time indices are denoted by $\mu,\nu,\ldots$ and
$\hat{\mu},\hat{\nu},\ldots$, respectively. Space-time signature is $(-++++++)$. Lower and upper $M,N=1,2,\ldots,5$ indices describe the fundamental and anti-fundamental representations $\mathbf{5}$ and $\bar{\mathbf{5}}$ of $SL(5)$ global symmetry, respectively. According to this convention, the ten vector fields $A^{MN}=A^{[MN]}$ transform as $\overline{\mathbf{10}}$ under $SL(5)$ while the two-forms $B_{M\mu\nu}$ transform in $\mathbf{5}$ representation. There are $14$ scalars living in $SL(5)/SO(5)$ coset and described by the coset representative ${\mc{V}_M}^A$, $A=1,2,\ldots, 5$. 
\\
\indent Fermionic fields, on the other hand, transform under the local $SO(5)\sim USp(4)$ R-symmetry. Indices $a,b,\ldots=1,2,3,4$ correspond to spinor representation of $SO(5)$ or equivalently the fundamental representation of $USp(4)$. The gravitini transform as $\mathbf{4}$ under $USp(4)$ while the spin-$\frac{1}{2}$ fields $\chi^{abc}$ transform as $\mathbf{16}$. The latter are subject to the conditions
\begin{equation}
\chi^{abc}=\chi^{[ab]c}\qquad \textrm{and}\qquad \Omega_{ab}\chi^{abc}=\chi^{[abc]}=0\, .
\end{equation} 
$\Omega_{ab}=\Omega_{[ab]}$ is $USp(4)$ symplectic form with the inverse $\Omega^{ab}=(\Omega_{ab})^*$ satisfying $\Omega_{ab}\Omega^{cb}=\delta^c_a$. Raising and lowering of $USp(4)$ indices $a,b,\ldots$ by $\Omega^{ab}$ and $\Omega_{ab}$ are related to complex conjugation for example
\begin{equation}
(V^{a})^*=\Omega_{ab}V^b\qquad \textrm{and}\qquad (V_a)^*=\Omega^{ab}V_b\, .
\end{equation}
All fermions are symplectic Majorana spinors subject to the conditions
\begin{equation}
\bar{\psi}^T_{\mu a}=\Omega_{ab}C\psi_\mu^b\qquad \textrm{and}\qquad \bar{\chi}^T_{abc}=\Omega_{ad}\Omega_{be}\Omega_{cf}C\chi^{def}
\end{equation}
where $C$ denotes the charge conjugation matrix obeying
\begin{equation}
C=C^T=-C^{-1}=-C^\dagger\, .
\end{equation}
The Dirac conjugate on a spinor $\Psi$ is defined by $\overline{\Psi}=\Psi^\dagger \gamma^0$. We will denote space-time gamma matrices by $\gamma^\mu$ as opposed to $\Gamma^\mu$ in the convention of \cite{N4_7D_Henning}.
\\
\indent The $SL(5)/SO(5)$ coset representative ${\mc{V}_M}^A$ transform under the global $SL(5)$ and local $SO(5)\sim USp(4)$ by left and right multiplications, respectively. Accordingly, the index $A$ can be described by an anti-symmetric pair of $USp(4)$ fundamental indices, and ${\mc{V}_M}^A$ can be written as ${\mc{V}_M}^{ab}$ subject to the condition
\begin{equation}
{\mc{V}_M}^{ab}\Omega_{ab}=0\, .
 \end{equation} 
The inverse of ${\mc{V}_M}^{ab}$ will be denoted by ${\mc{V}_{ab}}^M$. We then have the following relations 
\begin{equation}
{\mc{V}_M}^{ab}{\mc{V}_{ab}}^N=\delta^N_M\qquad \textrm{and}\qquad {\mc{V}_{ab}}^M{\mc{V}_M}^{cd}=\delta^{cd}_{ab}-\frac{1}{4}\Omega_{ab}\Omega^{cd}
\end{equation}
with $\delta^{cd}_{ab}=\delta^{[c}_a\delta^{d]}_b$. It should be noted that the $SL(5)/SO(5)$ coset can also be described by a unimodular symmetric matrix $\mc{M}_{MN}$ defined by
\begin{equation}
\mc{M}_{MN}={\mc{V}_M}^{ab}{\mc{V}_N}^{cd}\Omega_{ac}\Omega_{bd}
\end{equation}
with its inverse given by $\mc{M}^{MN}={\mc{V}_{ab}}^M{\mc{V}_{cd}}^N\Omega^{ac}\Omega^{bd}$.
\\
\indent The most general gaugings of $N=4$ supergravity can be efficiently described by using the embedding tensor ${\Theta_{MN,P}}^Q$. This tensor describes an embedding of a gauge group $G_0$ in the global symmetry group $SL(5)$ via the covariant derivative
\begin{equation} 
D_\mu=\nabla_\mu-gA^{MN}_\mu {\Theta_{MN,P}}^Q {t^P}_Q\label{gauge_covariant_derivative}
\end{equation}  
with $\nabla_\mu$ being the space-time covariant derivative including (possibly) composite $SO(5)$ connections. ${t^P}_Q$ are $SL(5)$ generators and $g$ is the gauge coupling constant.
\\
\indent The covariant derivative implies that the embedding tensor lives in the product representation between the conjugate representation of the vector fields and the adjoint representation of $SL(5)$ 
\begin{equation}
\mathbf{10}\otimes \mathbf{24}=\mathbf{10}+\mathbf{15}+\overline{\mathbf{40}}+\mathbf{175}. 
\end{equation}
Among the resulting irreducible representations, supersymmetry allows only the embedding tensor in $\mathbf{15}$ and $\overline{\mathbf{40}}$. These representations will be denoted respectively by $Y_{MN}$ and $Z^{MN,P}$ with $Y_{MN}=Y_{(MN)}$, $Z^{MN,P}=Z^{[MN],P}$ and $Z^{[MN,P]}=0$. In terms of $Y_{MN}$ and $Z^{MN,P}$, the embedding tensor can be written as      
\begin{equation}\label{embedding tensor}
{\Theta_{MN,P}}^Q=\delta^Q_{[M}Y_{N]P}-2\epsilon_{MNPRS}Z^{RS,Q}\, .
\end{equation}      
\indent In order to define a viable gauging, the embedding tensor needs to satsisfy the so-called quadratic constraint to ensure that the gauge generators $X_{MN}={\Theta_{MN,P}}^Q{t^P}_Q$ form a closed subalgebra of $SL(5)$
\begin{equation}
\left[X_{MN},X_{PQ}\right]=-{(X_{MN})_{PQ}}^{RS}X_{RS}.\label{quadratic_constraint}
\end{equation}
In the fundamental representation $\mathbf{5}$ of $SL(5)$, gauge generators ${(X_{MN})_P}^Q$ can be written as
\begin{equation}\label{OurgaugeGen}
{(X_{MN})_P}^Q= {\Theta_{MN,P}}^Q=\delta^Q_{[M}Y_{N]P}-2\epsilon_{MNPRS}Z^{RS,Q}
\end{equation}
while in the $\mathbf{10}$ representation, these generators are given by
\begin{equation}\label{10gaugeGen}
{(X_{MN})_{PQ}}^{RS}=2{X_{MN,[P}}^{[R}\delta^{S]}_{Q]}\, .
\end{equation}
In terms of $Y_{MN}$ and $Z^{MN,P}$, the quadratic constraint \eqref{quadratic_constraint} reads
\begin{equation}\label{QuadCon} 
Y_{MQ}Z^{QN,P}+2\epsilon_{MRSTU}Z^{RS,N}Z^{TU,P}=0\, .
\end{equation}
It should also be noted that this constraint implies that the embedding tensor is gauge invariant
\begin{equation}
{(X_{MN})_{PQ}}^{TU}{\Theta_{TU,R}}^S+{(X_{MN})_R}^{T}{\Theta_{PQ,T}}^S-{(X_{MN})_{T}}^{S}{\Theta_{PQ,R}}^T=0\, .
\end{equation}
\indent The introduction of the minimal coupling \eqref{gauge_covariant_derivative} usually breaks the original supersymmetry. To restore supersymmetry, modifications to the Lagrangian and supersymmetry transformations are needed. In addition to the introduction of fermionic mass-like terms and scalar potential, gaugings also lead to hierarchies of non-abelian vector and tensor fields of different ranks. In this paper, we are interested only in domain wall solutions with only the metric and scalars non-vanishing. We will set all vector and tensor fields to zero from now on. It is straightforward to verify that for all solutions under consideration here, this is indeed a consistent truncation. 
\\
\indent The bosonic Lagrangian with only the metric and scalar fields reads
\begin{equation}
e^{-1}\mathcal{L}=\frac{1}{2}R-\frac{1}{2}{P_{\mu ab}}^{cd}{{P^\mu}_{cd}}^{ab}-V,\label{boconic_L}
\end{equation}
and the supersymmetry transformations of $\psi^a_\mu$ and $\chi^{abc}$ are given by
\begin{eqnarray}
\delta\psi^a_\mu&=& D_\mu\epsilon^a-g\gamma_\mu A^{ab}_1\Omega_{bc}\epsilon^c,\\
\delta\chi^{abc}&=& 2\Omega^{cd}{P_{\mu de}}^{ab}\gamma^\mu\epsilon^e+gA^{d,abc}_2\Omega_{de}\epsilon^e\, .
\end{eqnarray}
\indent The covariant derivative of $\epsilon^a$ is defined as
\begin{equation}
D_\mu\epsilon^a= \pd_\mu\epsilon^a+\frac{1}{4}{\omega_\mu}^{\hat{\nu}\hat{\rho}}\gamma_{\hat{\nu}\hat{\rho}}\epsilon^a-{Q_{\mu b}}^a\epsilon^b\, .
\end{equation}
The vielbein on the $SL(5)/SO(5)$ coset ${P_{\mu ab}}^{cd}$ and the $SO(5)\sim USp(4)$ composite connection ${Q_{\mu a}}^b$ are obtained from the relation 
\begin{equation}
{\mathcal{V}_{ab}}^M\partial_\mu{\mathcal{V}_M}^{cd}={P_{\mu ab}}^{cd}+2{Q_{\mu [a}}^{[c}\delta^{d]}_{b]}\, .
\end{equation}
\indent $A_1$ and $A_2$ tensors are given in terms of scalar fields and the embedding tensor 
\begin{eqnarray}
A^{ab}_1&=&-\frac{1}{4\sqrt{2}}\left(\frac{1}{4}B\Omega^{ab}+\frac{1}{5}C^{ab}\right),\\
A^{d,abc}_2&=&\frac{1}{2\sqrt{2}}\left[\Omega^{ec}\Omega^{fd}({C^{ab}}_{ef}-{B^{ab}}_{ef})\phantom{\frac{1}{4}}\right. \nonumber \\
& &\left. +\frac{1}{4}\left(C^{ab}\Omega^{cd}+\frac{1}{5}\Omega^{ab}C^{cd}+\frac{4}{5}\Omega^{c[a}C^{b]d}\right)\right]
\end{eqnarray}
in which $B$ and $C$ tensors are defined by
\begin{eqnarray}
B&=&\ \frac{\sqrt{2}}{5}\Omega^{ac}\Omega^{bd}Y_{ab,cd},\\
{B^{ab}}_{cd}&=&\ \sqrt{2}\left[\Omega^{ae}\Omega^{bf}\delta^{gh}_{cd}-\frac{1}{5}\left(\delta^{ab}_{cd}-\frac{1}{4}\Omega^{ab}\Omega_{cd}\right)\Omega^{eg}\Omega^{fh}\right]Y_{ef,gh},\\
C^{ab}&=&\ 8\Omega_{cd}Z^{(ac)[bd]},\\
{C^{ab}}_{cd}&=&\ 8\left(-\Omega_{ce}\Omega_{df}\delta^{ab}_{gh}+\Omega_{g(c}\delta^{ab}_{d)e}\Omega_{fh}\right)Z^{(ef)[gh]}
\end{eqnarray}
with
\begin{eqnarray}
Y_{ab,cd}&=&{\mathcal{V}_{ab}}^M{\mathcal{V}_{cd}}^NY_{MN},\\
Z^{(ac)[ef]}&=&\sqrt{2}{\mathcal{V}_M}^{ab}{\mathcal{V}_N}^{cd}{\mathcal{V}_P}^{ef}\Omega_{bd}Z^{MN,P}\, .
\end{eqnarray}
Finally, the scalar potential is given by
\begin{equation}
V=-15A_1^{ab}A_{1 ab}+ \frac{1}{8}A_2^{a,bcd}A_{2 a,bcd}=-15|A_1|^2+\frac{1}{8}|A_2|^2\, .
\end{equation}
\indent It should also be noted that the Lagrangian \eqref{boconic_L} can be written in a $USp(4)$ invariant form as
\begin{equation}
e^{-1}\mathcal{L}=\frac{1}{2}R+\frac{1}{8}(\pd_\mu\mathcal{M}_{MN})(\pd^\mu\mathcal{M}^{MN})-V\label{7D_Lgauge}
\end{equation}
with the scalar potential given by
\begin{eqnarray}
V&=& \frac{g^2}{64}\left[2\mathcal{M}^{MN}Y_{NP}\mathcal{M}^{PQ}Y_{QM}-(\mathcal{M}^{MN}Y_{MN})^2\right]\nonumber \\
& &+g^2Z^{MN,P}Z^{QR,S}\left(\mathcal{M}_{MQ}\mathcal{M}_{NR}\mathcal{M}_{PS}
-\mathcal{M}_{MQ}\mathcal{M}_{NP}\mathcal{M}_{RS}\right).
\end{eqnarray}

%%%%%%%%%%%%%%%%%%%%%%%%%%%%%%%%%%%%%%%%%%%%%%%%%%%%%%%%%%%%%%%%%%%%%%%%%%%%%%%%%%%%%%%%%%%%%%%%%%%%%%%%%%%%%%%%%%%%%%%%%%%%%%%%%%%%%%%%%
\section{Supersymmetric domain walls from gaugings in $\mathbf{15}$ representation}\label{Y_gauging}
In this section, we consider gauge groups arising from the embedding tensor in $\mathbf{15}$ representation. It is readily seen from \eqref{QuadCon} that setting $Z^{MN,P}=0$ trivially satisfies the quadratic constraint. Therefore, any symmetric tensor $Y_{MN}$ leads to an admissible gauge group. The $SL(5)$ symmetry can be used to fix the form of $Y_{MN}$ to be
\begin{equation}
Y_{MN}=\text{diag}(\underbrace{1,..,1}_p,\underbrace{-1,..,-1}_q,\underbrace{0,..,0}_r),\qquad p+q+r=5\, .
\end{equation}
The corresponding gauge generators are given by
\begin{equation}
{(X_{MN})_P}^Q=\delta^Q_{[M}Y_{N]P}
\end{equation}
which give rise to the gauge group
\begin{equation}\label{CSOgaugegroup}
CSO(p,q,r)\sim SO(p,q)\ltimes\mathbf{R}^{(p+q)r}\, .
\end{equation}
\indent In order to find supersymmetric solutions, we restrict ourselves to a subset of scalars invariant under a certain symmetry group $H_0\subset G_0$ following the approach introduced in \cite{warner}. The metric takes the form of standard domain wall ansatz 
\begin{equation}\label{DWmetric}
ds_7^2=e^{2A(r)}\eta_{\alpha \beta}dx^\alpha dx^\beta+dr^2
\end{equation}
where $\alpha,\beta=0,1,..,5$ and $A(r)$ is a warp factor depending only on the radial coordinate $r$. 
 \\
 \indent Non-compact generators of $SL(5)$ are given by $5\times 5$ symmetric traceless matrices. To obtain an explicit parametrization of the coset representative ${\mc{V}_M}^A$, it is useful to introduce $GL(5)$ matrices
\begin{equation}\label{GL(5,R)}
{(e_{MN})_K}^L=\delta_{MK}\delta_{N}^L\, .
\end{equation}  
To convert the $SO(5)$ vector indices $A,B,\ldots =1,2,\ldots, 5$ to a pair of anti-symmetric $USp(4)$ indices $a,b,\ldots =1,2,3,4$, we use a convenient choice of $SO(5)$ gamma matrices given by 
\begin{eqnarray}
\Gamma_1&=&-\sigma_2\otimes\sigma_2, \qquad  \Gamma_2=\mathbf{I}_2\otimes\sigma_1,\qquad \Gamma_3=\mathbf{I}_2\otimes\sigma_3,\nonumber  \\ 
\Gamma_4&=&\sigma_1\otimes\sigma_2, \qquad \Gamma_5=\sigma_3\otimes\sigma_2
\end{eqnarray}
where $\sigma_i$ are the usual Pauli matrices. $\Gamma_A$ satisfy the following relations
\begin{eqnarray}
\{\Gamma_A,\Gamma_B\}&=&2\delta_{AB}\mathbf{I}_4,\qquad  (\Gamma_A)^{ab}=-(\Gamma_A)^{ba}, \nonumber \\ 
\Omega_{ab}(\Gamma_A)^{ab}&=&0, \qquad ((\Gamma_A)^{ab})^*=\Omega_{ac}\Omega_{bd}(\Gamma_A)^{cd}\, .
\end{eqnarray}
The symplectic form of $USp(4)$ is taken to be
\begin{equation}
\Omega_{ab}=\Omega^{ab}=\mathbf{I}_2\otimes i\sigma_2\, .
\end{equation} 
The coset representative of the form ${\mc{V}_M}^{ab}$ and the inverse ${\mc{V}_{ab}}^M$ are then obtained by using the relations
\begin{equation}
{\mathcal{V}_M}^{ab}=\frac{1}{2} {\mathcal{V}_M}^A(\Gamma_A)^{ab}\qquad \textrm{and}\qquad 
{\mathcal{V}_{ab}}^M=\frac{1}{2} {\mathcal{V}_A}^M(\Gamma^A)_{ab}\, .
\end{equation}
We are now in a position to set up BPS equations and look for domain wall solutions with different unbroken symmetries.

\subsection{$SO(4)$ symmetric domain walls}
We begin with a simple solution with $SO(4)$ symmetry. The gauge groups that contain $SO(4)$ as a subgroup are $SO(5)$, $SO(4,1)$ and $CSO(4,0,1)$. To incorporate all of these gauge groups within a single framework, we write the embedding tensor in the form
\begin{equation}\label{YSO(4)YMN}
Y_{MN}=\text{diag}(1,1,1,1,\kappa)
\end{equation}
with $\kappa=1,0,-1$ corresponding to SO(5), CSO(4,0,1), and SO(4,1) gauge groups, respectively. There is one $SO(4)$ singlet scalar corresponding to the non-compact generator
\begin{equation}\label{YSO(4)Ys}
\hat{Y}=e_{11}+e_{22}+e_{33}+e_{44}-4e_{55}\, .
\end{equation}
The coset representative can be written as
\begin{equation}
\mathcal{V}=e^{\phi\hat{Y}}\, .
\end{equation}
The scalar potential for this $SO(4)$ invariant scalar is given by
\begin{equation}\label{YSO(4)Pot}
V=-\frac{g^2}{64}e^{-4 \phi} (8+8\kappa e^{10 \phi}-\kappa^2e^{20 \phi}).
\end{equation}
It can be verified that, for $\kappa=1$, this potential admits two $AdS_7$ critical points at $\phi=0$ and $\phi=\frac{1}{10}\ln 2$. These critical points have already been studied in \cite{7DN4_gauged2}. The first critical point has $SO(5)$ symmetry and preserves all supersymmetry. Upon uplifting, this vacuum corresponds to $AdS_7\times S^4$ solutions of eleven-dimensional supergravity. The cosmological constant and $AdS_7$ radius are given by
\begin{equation}
V_0=-\frac{15}{64}g^2\qquad \textrm{and}\qquad L=\sqrt{-\frac{15}{V_0}}=\frac{8}{g}\, .
\end{equation}    
The second critical point is $SO(4)$ symmetric and breaks all supersymmetry. This non-supersymmetric $AdS_7$ vacuum is unstable \cite{7DN4_gauged2}.
\\
\indent In order to setup the corresponding BPS equations, we impose a projector 
\begin{equation}\label{pureYProj}
\gamma_r\epsilon^a=\epsilon^a
\end{equation}
and obtain the following BPS equations from $\delta \psi^a_\alpha=0$ and $\delta\chi^{abc}=0$ conditions
\begin{eqnarray}
A'&=& \frac{g}{40}e^{-2 \phi}(4 +\kappa e^{10 \phi}),\label{YSO(4)BPS1}\\
\phi'&=& \frac{g}{20}e^{-2 \phi}(1 - \kappa e^{10 \phi}).\label{YSO(4)BPS2}
\end{eqnarray}
The condition $\delta \psi^a_r=0$ gives the usual solution for the Killing spinors 
\begin{equation}
\epsilon^a=e^{\frac{A}{2}}\epsilon^a_0\label{DW_Killing_spinor}	
\end{equation}	
with the constant spinors $\epsilon^a_0$ satisfying $\gamma_r\epsilon^a_0=\epsilon^a_0$. The solution is then half-supersymmetric.
\\
\indent The above BPS equations can be readily solved to obtain the solution
\begin{eqnarray}
A&=&2\phi-\frac{1}{4}\ln (1-\kappa e^{10\phi}),\\
e^{5\phi}&=&\frac{1}{\sqrt{\kappa}}\tanh\left[\frac{\sqrt{\kappa}}{4}(g\rho+C)\right]\label{Y_DW_SO4}
\end{eqnarray}
with the new radial coordinate $\rho$ defined by $\frac{d\rho}{dr}=e^{3\phi}$. The integration constant $C$ can be removed by shifting the coordinate $\rho$. We have also neglected an additive integration constant for $A$ since it can be absorbed by rescaling the coordinates $x^\alpha$.
\\
\indent Note that for $\kappa=-1$, the solution for $\phi$ can be written as
\begin{equation}
e^{5\phi}=\tan\left[\frac{1}{4}(g\rho+C)\right].
\end{equation}  		
For $\kappa=0$, we find
\begin{equation}
e^{5\phi}=\frac{1}{4}(g\rho+C).
\end{equation}		
	
\subsection{$SO(3)\times SO(2)$ symmetric domain walls}
We now consider another residual symmetry $SO(3)\times SO(2)$ which is possible only for $SO(5)$ and $SO(3,2)$ gauge groups. In this case, we write the embedding tensor as
\begin{equation}
Y_{MN}=\textrm{diag}(1,1,1,\sigma,\sigma)
 \end{equation}
with $\sigma=1$ and $\sigma=-1$ corresponding to $SO(5)$ and $SO(3,2)$, respectively.
\\
\indent The only one $SO(3)\times SO(2)$ singlet scalar corresponds to the non-compact generator
\begin{equation}
\tilde{Y}=2e_{11}+2e_{22}+2e_{33}-3e_{44}-3e_{55}\, .
\end{equation}
With the coset representative
\begin{equation}
\mc{V}=e^{\phi\tilde{Y}},
\end{equation}
we find the scalar potential
\begin{equation}
V=-\frac{3}{64}g^2e^{-8\phi}(1+4\sigma e^{10\phi})
\end{equation}
which admits an $AdS_7$ critical point at $\phi=0$ for $\sigma=1$.   
\\
\indent The BPS equations are given by
\begin{eqnarray}
\phi'&=&-\frac{1}{20}ge^{-4\phi}(\sigma e^{10\phi}-1),\\
A'&=&\frac{1}{40}ge^{-4\phi}(3+2\sigma e^{10\phi}).
\end{eqnarray}        
By defining a new radial coordinate $\rho$ by the relation $\frac{d\rho}{dr}=e^{\phi}$, we obtain the solution
\begin{eqnarray}
A&=&\frac{3}{2}\phi-\frac{1}{4}\ln (1-\sigma e^{10\phi}),\\
e^{5\phi}&=&\frac{1}{\sqrt{\sigma}}\tanh\left[\frac{\sqrt{\sigma}}{4}(g\rho+C)\right].
\end{eqnarray}
This solution is very similar to the $SO(4)$ symmetric solution. 

\subsection{$SO(3)$ symmetric domain walls}\label{SO3_DW_Y}
When the residual symmetry of the solutions is smaller, we find more interesting solutions. We now consider domain wall solutions with $SO(3)$ symmetry. There are many gauge groups containing $SO(3)$ subgroup with the embedding tensor given by
\begin{equation}
Y_{MN}=\text{diag}(1,1,1,\sigma,\kappa).
\end{equation}
There are three scalar singlets under $SO(3)$ symmetry generated by gauge generators $X_{MN}$, $M,N=1,2,3$. These singlets correspond to the following non-compact generators of $SL(5)$
\begin{eqnarray}\label{YSO(3)Ys}
\hat{Y}_1&=& 2e_{1,1}+2e_{2,2}+2e_{3,3}-3e_{4,4}-3e_{5,5},\nonumber \\
\hat{Y}_2&=& e_{4,5}+e_{5,4},\nonumber \\
\hat{Y}_3&=& e_{4,4}-e_{5,5}\, .
\end{eqnarray}  
Using the parametrization of the coset representative
\begin{equation}\label{YSO(3)Ys}
\mathcal{V}=e^{\phi_1\hat{Y}_1+\phi_2\hat{Y}_2+\phi_3\hat{Y}_3},
\end{equation}
we obtain the scalar potential
\begin{eqnarray}\label{YSO(3)Pot}
V&=&-\frac{g^2}{64}\left[3e^{-8\phi_1}+6e^{2\phi_1}\left[(\kappa+\sigma)\cosh{2\phi_2}\cosh{2\phi_3}+(\kappa-\sigma)\sinh{2\phi_3}\right]\right. \nonumber \\
& &+\frac{1}{4}e^{12\phi_1}\left[\kappa^2+10\kappa\sigma+\sigma^2-(3\kappa^2-2\kappa\sigma+3\sigma^2)\cosh{4\phi_3}\right. \nonumber \\
& &\left.\left.-(\kappa+\sigma)^2\cosh{4\phi_2}(1+\cosh{4\phi_3})-4(\kappa^2-\sigma^2)\cosh{2\phi_2}\sinh{4\phi_3}\right]\right].\quad
\end{eqnarray}
\indent This potential admits two $AdS_7$ critical point for $\kappa=\sigma=1$. The first one is at $\phi_1=\phi_2=\phi_3=0$ corresponding to the $N=4$ supersymmetric $AdS_7$ with $SO(5)$ symmetry. Another critical point is non-supersymmetric and given by
\begin{eqnarray}
\phi_1&=&\frac{1}{20}\ln{2},\qquad \phi_2=\frac{1}{4}\ln{2},\nonumber \\ 
\phi_3&=&0,\qquad V_0=-\frac{5g^2}{16\times2^{2/5}}\, . 
\end{eqnarray}
It should also be noted that for $\phi_2=5\phi_1$, the residual symmetry is enhanced to $SO(4)$. As a check, we can compute all scalar masses at this critical point. The result is 
\begin{equation}
m^2L^2=(12,0_{\times 4},-12_{\times 9}),\qquad L=\frac{2^{\frac{11}{5}}\sqrt{3}}{g}
\end{equation}
which contains the value $m^2L^2$ that violates the BF bound $m^2L^2=-9$. Therefore, this critical point is unstable as already shown in \cite{7DN4_gauged2}. The four Goldstone bosons corresponding to the broken generators $X_{a4}-X_{a5}$, $a=1,2,3$ and $X_{45}$.
\\
\indent Using the same procedure as in the previous cases, we find the following BPS equation
\begin{eqnarray}\label{YSO(3)BPS}
A'&=& \frac{g}{40}e^{-4\phi_1}\left[3+e^{10\phi_1}\left[(\kappa+\sigma)\cosh{2\phi_2}\cosh{2\phi_3}+(\kappa-\sigma)\sinh{2\phi_3}\right]\right],\\
\phi'_1&=&\ \frac{g}{40}e^{-4\phi_1}\left[2-e^{10\phi_1}[(\kappa+\sigma)\cosh{2\phi_2}\cosh{2\phi_3}-(\kappa-\sigma)\sinh{2\phi_3}]\right],\\
\phi'_2&=&\ -\frac{g}{8}e^{6\phi_1}(\kappa+\sigma)\sinh{2\phi_2}\text{\ sech} 2\phi_3,\\
\phi'_3&=&\ -\frac{g}{8}e^{6\phi_1}\left[(\kappa-\sigma)\cosh{2\phi_3}+(\kappa+\sigma)\cosh{2\phi_2}\sinh{2\phi_3}\right].
\end{eqnarray}
To find explicit solutions, it is useful to separately discuss various possible values of $\kappa$ and $\sigma$.

\subsubsection{Domain walls in $CSO(3,0,2)$ gauge group} 
We begin with the simplest case for $\sigma=\kappa=0$ corresponding to a non-semisimple $CSO(3,0,2)$ gauge group. In this case, we find $\phi'_2=\phi'_3=0$. Furthermore, it can be checked that $\frac{\pd V}{\pd\phi_2}=\frac{\pd V}{\pd \phi_3}=0$ at $\phi_2=\phi_3=0$. Therefore, scalars $\phi_2$ and $\phi_3$ can be consistently truncated out.
\\
\indent After setting $\phi_2=\phi_3=0$, we find a domain wall solution
\begin{equation}
\phi_1=\frac{1}{4}\ln \left[\frac{gr}{5}+C\right]\qquad \textrm{and}\qquad A=\frac{3}{8}\ln\left[\frac{gr}{5}+C\right].
\end{equation}

\subsubsection{Domain walls in $CSO(4,0,1)$ and $CSO(3,1,1)$ gauge groups}
For $\kappa=0$ and $\sigma\neq 0$, the gauge group is either $CSO(4,0,1)$ or $CSO(3,1,1)$ depending on the value of $\sigma=1$ or $\sigma=-1$. Using a new radial coordinate $\rho$ defined by $\frac{d\rho}{dr}=e^{6\phi_1}$, a domain wall solution to the BPS equations can be found 
\begin{eqnarray}
\phi_2&=&\frac{1}{4}\ln\left[\frac{g^2\rho^2+(C_2-8)^2}{g^2\rho^2+C_2^2}\right],\label{SO3_DW_CSO41_1}\\
\phi_3&=&\frac{1}{4}\ln\left[\frac{e^{2\phi_2}-e^{4\phi_2+C_3}+e^{C_3}+1}{e^{2\phi_2}+e^{4\phi_2+C_3}-e^{C_3}-1}\right],\label{SO3_DW_CSO41_2}\\
\phi_1&=&\frac{1}{10}\ln\left[\frac{2(e^{C_1}-e^{4\phi_2+C_1}-1)}{\sigma\sqrt{(e^{4\phi_2}-1)(1+2e^{C_3}+e^{2C_3}-e^{4\phi_2+2C_3})}}\right],\label{SO3_DW_CSO41_3}\\
A&=&-\phi_1-\ln(e^{4\phi_2}-1)+\ln(e^{C_1}-e^{4\phi_2+C_1}-1).\label{SO3_DW_CSO41_4}
\end{eqnarray}

\subsubsection{Domain walls in $SO(4,1)$ gauge group}
In this case, $\sigma=-\kappa=1$, we find that $\phi_2'=0$. It can also be checked that $\phi_2$ can be consistently truncated out. Note also that the corresponding non-compact generator $\hat{Y}_2$ is one of the non-compact generators of $SO(4,1)$, namely $X_{45}$. $\phi_2$ is then identified with a Goldstone boson of the symmetry breaking $SO(4,1)\rightarrow SO(4)\rightarrow SO(3)$ at the vacuum.
\\
\indent Taking $\phi_2=0$ and redefining the radial coordinate $r$ to $\rho$ via $\frac{d\rho}{dr}=e^{6\phi_1}$, we obtain a domain wall solution 
\begin{eqnarray}
e^{2\phi_3}&=&\tan\left[\frac{g\rho}{4}+C_3\right],\\
\phi_1&=&-\frac{1}{5}\phi_3+\frac{1}{10}\ln\left[C_1(1+e^{4\phi_3})-1\right],\\
A&=&\frac{1}{5}\phi_3-\frac{1}{4}\ln(1+e^{4\phi_3})+\frac{3}{20}\ln\left[C_1(1+e^{4\phi_3})-1\right].
\end{eqnarray}

\subsubsection{Domain walls in $SO(5)$ and $SO(3,2)$ gauge groups}
We now look at the last possibility $\kappa=\sigma=\pm 1$ corresponding to $SO(5)$ and $SO(3,2)$ gauge groups. In term of the new radial coordinate $\rho$ as defined in the previous cases, we find a domain wall solution
\begin{eqnarray}
\phi_2&=&\frac{1}{4}\ln \left[\frac{1+e^{g\sigma\rho}+4e^{g\sigma\rho+2C_3}-2e^{\frac{1}{2}g\sigma\rho}}
{1+e^{g\sigma\rho}+4e^{g\sigma\rho+2C_3}+2e^{\frac{1}{2}g\sigma\rho}}\right],\\
\phi_3&=&\frac{1}{4}\ln\left[\frac{e^{2\phi_2}+e^{4\phi_2+C_3}-e^{C_3}}{e^{2\phi_2}-e^{4\phi_2+C_3}+e^{C_3}}\right],\\
\phi_1&=&\frac{1}{10}\ln\left[\sigma\left[1+C_1(e^{4\phi_2}-1)\right]\sqrt{e^{8\phi_2+2C_3}+e^{2C_3}-e^{4\phi_2}-2e^{4\phi_2+2C_3}}\right],\\
A&=&-\phi_1+\frac{1}{4}\ln(e^{4\phi_2}-1)-\frac{1}{4}\ln[1+C_1(e^{4\phi_2}-1)].
\end{eqnarray}

\subsection{$SO(2)\times SO(2)$ symmetric domain walls}\label{SO2_SO2_DW_Y}
We consider another truncation to $SO(2)\times SO(2)$ invariant scalars corresponding to $SL(5)$ non-compact generators
\begin{eqnarray}
\tilde{Y}_1&=& e_{11}+e_{22}-2e_{55},\\
\tilde{Y}_2&=&e_{33}+e_{44}-2e_{55}\, .
\end{eqnarray}
In this case, the embedding tensor takes the form of
\begin{equation}
Y_{MN}=\textrm{diag}(1,1,\sigma,\sigma,\kappa)
\end{equation} 
which encodes various possible gauge groups depending on the values of $\sigma$ and $\kappa$. These gauge groups are $SO(5)$ ($\sigma=\kappa=1$), $SO(4,1)$ ($\sigma=-\kappa=1$), $SO(3,2)$ ($\sigma=-\kappa=-1$), $CSO(4,0,1)$ ($\sigma=1,\kappa=0$) and $CSO(2,2,1)$ ($\sigma=-1,\kappa=0$).
\\
\indent With the parametrization of the coset representative
\begin{equation}\label{YSO(2)SO2Ys}
\mathcal{V}=e^{\phi_1\tilde{Y}_1+\phi_2\tilde{Y}_2},
\end{equation}
we find the scalar potential
\begin{equation}\label{YSO(2)diagPot}
V=-\frac{1}{64}g^2e^{-2(\phi_1+\phi_2)}\left[8\sigma -\kappa^2e^{10(\phi_1+\phi_2)}+4\kappa(e^{4\phi_1+6\phi_2}+\sigma e^{6\phi_1+4\phi_2})\right].
\end{equation}
For $SO(5)$ gauge group, there are two $AdS_7$ critical points at $\phi_1=\phi_2=0$ and $\phi_1=\phi_2=\frac{1}{10}\ln 2$. The former is, as in other cases, the $N=4$ supersymmetric one while the latter is a non-supersymmetric critical point. Note also that this non-supersymmetric $AdS_7$ has $SO(4)$ symmetry since the $SO(2)\times SO(2)$ symmetry is enhanced to $SO(4)$ when $\phi_1=\phi_2$. This critical point is unstable as previously mentioned.
\\
\indent The BPS equations in this case read
\begin{eqnarray}
A'&=& \frac{g}{40}\left[2e^{-2\phi_1}+2\sigma e^{-2\phi_2}+\kappa e^{4(\phi_1+\phi_2)}\right],\\
\phi_1'&=& \frac{g}{20}\left[3e^{-2\phi_1}-\kappa e^{4(\phi_1+\phi_2)}-2\sigma e^{-2\phi_2}\right],\\
\phi_2'&=& \frac{g}{20}\left[3\sigma e^{-2\phi_2}-2e^{-2\phi_1}-\kappa e^{4(\phi_1+\phi_2)}\right].
\end{eqnarray}
Defining a new radial coordinate $\rho$ by $\frac{d\rho}{dr}=e^{-2\phi_1}$, we find a domain wall solution
\begin{eqnarray}
\phi_2&=&-\frac{3}{2}\phi_1-\frac{1}{4}\ln\left[\kappa-\kappa e^{C_2-\frac{g\rho}{2}}\right],\label{SO2_SO2_YDW1}\\
\phi_1&=&-\frac{1}{10}\ln\left[\kappa-\kappa e^{C_1-\frac{g\rho}{2}}\right]-\frac{1}{5}\ln\left[\sigma-\sigma e^{C_2-\frac{g\rho}{2}}\right],\label{SO2_SO2_YDW2}\\
A&=&\frac{g\rho}{8}+\frac{1}{10}\ln \left[1-e^{C_1-\frac{g\rho}{2}}\right]+\frac{1}{20}\ln \left[1-e^{C_2-\frac{g\rho}{2}}\right].\label{SO2_SO2_YDW3}
\end{eqnarray}

\subsection{Uplift to eleven dimensions and holographic RG flows}
For $SO(5)$ gauge group, the seven-dimensional gauged supergravity can be obtained from a consistent truncation of eleven-dimensional supergravity on $S^4$. Therefore, the domain wall solutions obtained previously can be uplifted to solutions of eleven-dimensional supergravity. Furthermore, these solutions are asymptotic to the $N=4$ supersymmetric $AdS_7$ vacuum corresponding to $N=(2,0)$ SCFT in six dimensions. According to the AdS/CFT correspondence, the domain walls can then be interpreted as holographic RG flows from six-dimensional $N=(2,0)$ SCFT to non-conformal field theories in the IR, see for example \cite{non-CFT_flow_Zaffaroni,Gubser_non_CFT}. We will consider this type of solutions including the uplift to eleven dimensions.

\subsubsection{RG flow preserving $SO(4)$ symmetry}    
We first consider a simple solution with $SO(4)$ symmetry. For $SO(5)$ gauge group, the domain wall solution reads
\begin{eqnarray}
\phi&=&\frac{1}{5}\ln \left[\frac{1-e^{\frac{1}{2}(C-g\rho)}}{1+e^{\frac{1}{2}(C-g\rho)}}\right],\\
A&=&2\phi-\frac{1}{4}\ln\left(1-e^{10\phi}\right),
\end{eqnarray}
with $\frac{d\rho}{dr}=e^{3\phi}$.
\\
\indent As $\rho\rightarrow \infty$, we find $\phi\rightarrow 0$ and $\rho\sim r$ with an asymptotic behavior
\begin{equation}
\phi\sim e^{-\frac{1}{2}gr}\sim e^{-\frac{4r}{L}}\qquad \textrm{and} \qquad A\sim \frac{1}{8}gr\sim \frac{r}{L},\qquad L=\frac{8}{g}\label{UV_behavior_SO4}
\end{equation}
which indicates that the solution approaches the supersymmetric $N=4$ $AdS_7$ critical point. The scalar $\phi$ is dual to an operator of dimension $\Delta=4$. Indeed, all scalars of the $N=4$ gauged supergravity are dual to operators of dimension four since they have the same mass with $m^2L^2=-8$.
\\
\indent As $g\rho\rightarrow C$, the solution is singular with the following behavior
\begin{equation}
\phi\sim \frac{1}{5}\ln (g\rho-C)\qquad \textrm{and}\qquad A\sim 2\phi\sim \frac{2}{5}\ln(g\rho-C)\, . 
\end{equation} 
We can now check that the scalar potential is bounded above with $V\rightarrow -\infty$ as $\phi\rightarrow -\infty$. This implies that the singularity is physically acceptable according to the criterion of \cite{Gubser_singularity}. In addition, we can use the truncation ansatz, reviewed in the appendix, to uplift this solution to eleven dimensions. 
\\
\indent With the parametrization of the $SL(5)/SO(5)$ coset
\begin{equation}
\mc{M}_{MN}=\textrm{diag}(e^{-8\phi},e^{2\phi},e^{2\phi},e^{2\phi},e^{2\phi})
\end{equation}
and the coordinates on $S^4$
\begin{equation}
\mu^M=(\mu^0,\mu^i)=(\cos\xi,\sin\xi \hat{\mu}^i),\qquad i=1,2,3,4
\end{equation}
with $\hat{\mu}^i$ being coordinates on $S^3$ satisfying $\hat{\mu}^i\hat{\mu}^i=1$, we find the eleven-dimensional metric and four-form field strength tensor
\begin{eqnarray}
d\hat{s}^2_{11}&=&\Delta^{\frac{1}{3}}\left(e^{2A}dx^2_{1,5}+dr^2\right)+\frac{16}{g^2}\Delta^{-\frac{2}{3}}\left[e^{-8\phi}\sin^2\xi d\xi^2\right.\nonumber \\
& &\left.+e^{2\phi}(\cos^2\xi d\xi^2+\sin^2\xi d\Omega^2_{(3)})\right],\\
\hat{F}_{(4)}&=&\frac{64}{g^3}\Delta^{-2}\sin^4\xi\left(U\sin\xi d\xi -10e^{6\phi}\phi'\cos\xi dr\right)\wedge \epsilon_{(3)}
\end{eqnarray}
with $d\Omega^2_{(3)}$ being the metric on a unit $S^3$ and
\begin{eqnarray}
\Delta&=&e^{8\phi}\cos^2\xi+e^{-2\phi}\sin^2\xi,\qquad \epsilon_{(3)}=\frac{1}{3!}\epsilon_{ijkl}\hat{\mu}^id\hat{\mu}^j\wedge d\hat{\mu}^k\wedge d\hat{\mu}^l,\nonumber \\
U&=&(e^{16\phi}-4e^{6\phi})\cos^2\xi-(e^{6\phi}+2e^{-4\phi})\sin^2\xi\, .
\end{eqnarray}
We see that the internal $S^4$ is deformed in such a way that an $S^3$ inside the $S^4$ is unchanged. The isometry of this $S^3$ is the $SO(4)$ residual symmetry of the seven-dimensional solution.
\\
\indent With the uplifted solution, we can look at the behavior of the metric component $\hat{g}_{00}=e^{2A}\Delta^{\frac{1}{3}}$ near the IR singularity. A straightforward computation gives   
\begin{equation}
\hat{g}_{00}\sim e^{\frac{10}{3}\phi}\rightarrow 0
\end{equation}
which means the singularity is physical according to the criterion given in \cite{maldacena_nogo}. This solution then describes an RG flow from six-dimensional $N=(2,0)$ SCFT to a non-conformal field theory in the IR. 
With the appearance of the normalizable mode in \eqref{UV_behavior_SO4}, the flow is driven by a vacuum expectation value of an operator of dimension $\Delta=4$ that breaks conformal symmetry and preserves only $SO(4)\subset SO(5)$ R-symmetry. It should be noted that this holographic RG flow has also been studied in \cite{Bobev_6D_holography} in the context of a truncation to half-maximal $N=2$ gauged supergravity.   

\subsubsection{RG flow preserving $SO(3)\times SO(2)$ symmetry}    
In this case, the flow solution reads
\begin{eqnarray}
\phi&=&\frac{1}{5}\ln\left[\frac{1-e^{-\frac{1}{2}(g\rho-C)}}{1+e^{-\frac{1}{2}(g\rho-C)}}\right],\\
A&=&\frac{3}{2}\phi-\frac{1}{4}\ln(1-e^{10\phi}),
\end{eqnarray}
with $\frac{d\rho}{dr}=e^\phi$.
\\
\indent As $r\rightarrow \infty$, we find 
\begin{equation}
\phi\sim e^{-\frac{4r}{L}}
\end{equation}
as in the previous case. As $g\rho\rightarrow C$, the solution becomes
\begin{eqnarray}
\phi\sim \frac{1}{5}\ln(g\rho-C)\qquad \textrm{and}\qquad A\sim \frac{3}{2}\phi\sim \frac{3}{10}\ln(g\rho-C).
\end{eqnarray}
Near the singularity, we find that the scalar potential is bounded above $V\rightarrow -\infty$. 
\\
\indent The uplifted solution can be obtained by using the $S^4$ coordinates
\begin{equation}
\mu^M=(\sin\xi \hat{\mu}^a,\cos\xi \cos\alpha,\cos\xi\sin \alpha),\qquad a=1,2,3
\end{equation}
with $\hat{\mu}^a\hat{\mu}^a=1$ and the scalar matrix
\begin{equation}
\mc{M}_{MN}=\textrm{diag}(e^{4\phi},e^{4\phi},e^{4\phi},e^{-6\phi},e^{-6\phi}).
\end{equation}
We find the eleven-dimensional solution
\begin{eqnarray}
d\hat{s}^2_{11}&=&\Delta^{\frac{1}{3}}\left(e^{2A}dx^2_{1,5}+dr^2\right)+\frac{16}{g^2}\left[e^{-6\phi}\cos^2\xi d\alpha^2+(e^{4\phi}\cos^2\xi+e^{-6\phi}\sin^2\xi)d\xi^2\right. \nonumber \\
& &\left.+e^{4\phi}\sin^2\xi d\hat{\mu}^ad\hat{\mu}^a\right],\\
\hat{F}_{(4)}&=&\frac{64}{3g^3}U\sin^3\xi\cos\xi \Delta^{-2}\left(\sin\xi d\xi+2e^{2\phi}\cos\xi \phi'dr\right)\wedge d\alpha\wedge \epsilon_{(2)}
\end{eqnarray}
where
\begin{equation}
\epsilon_{(2)}=\frac{1}{2}\epsilon_{abc}\hat{\mu}^ad\hat{\mu}^b\wedge d\hat{\mu}^c\, .
\end{equation}
We can see that the unbroken $SO(3)\times SO(2)$ symmetry corresponds to the isometry of the $S^2$, with the metric $d\Omega^2_{(2)}=d\hat{\mu}^ad\hat{\mu}^a$, inside the $S^4$ and the isometry of the $S^1$ parametrized by the coordinate $\alpha$.
\\
\indent From the eleven-dimensional metric, we find
\begin{equation}
\hat{g}_{00}\sim e^{\frac{5}{3}\phi}\rightarrow 0\, .
\end{equation}
The singularity is accordingly physical \cite{maldacena_nogo}, and the full solution describes an RG flow from $N=(2,0)$ SCFT to a non-conformal field theory with $SO(3)\times SO(2)$ symmetry. 

\subsubsection{RG flow preserving $SO(2)\times SO(2)$ symmetry}    
The flow solution is given by 
\begin{eqnarray}
\phi_1&=&-\frac{1}{10}\ln (1-e^{C_1-\frac{g\rho}{2}})-\frac{1}{5}\ln (1-e^{C_2-\frac{g\rho}{2}}),\\
\phi_2&=&-\frac{3}{2}\phi_1-\frac{1}{4}\ln(1-e^{C_1-\frac{g\rho}{2}}),\\
A&=&\frac{1}{8}g\rho+\frac{1}{20}\ln (1-e^{C_1-\frac{g\rho}{2}})+\frac{1}{10}\ln (1-e^{C_2-\frac{g\rho}{2}})
\end{eqnarray}
with $\frac{d\rho}{dr}=e^{-2\phi_1}$.
\\
\indent We can perform an uplift by using
\begin{eqnarray}
\mc{M}_{MN}&=&\textrm{diag}(e^{-4(\phi_1+\phi_2)},e^{2\phi_1},e^{2\phi_1},e^{2\phi_2},e^{2\phi_2}),\nonumber \\
\mu^M&=&(\cos\xi,\sin\xi\cos\psi\cos\alpha,\sin\xi\cos\psi\sin\alpha,\nonumber \\
& &\sin\xi\sin\psi\cos\beta,\sin\xi\sin\psi\sin\beta).
\end{eqnarray}
The corresponding eleven-dimensional metric is given by
\begin{eqnarray}
d\hat{s}^2_{11}&=&\Delta^{\frac{1}{3}}\left(e^{2A}dx^2_{1,5}+dr^2\right)+\frac{16}{g^2}\Delta^{-\frac{2}{3}}\left[e^{-4(\phi_1+\phi_2)}\sin^2\xi d\xi^2 \right.\nonumber \\
& &+e^{2\phi_1}\left(\cos^2\xi\cos^2\psi d\xi^2+\sin^2\xi\sin^2\psi d\psi^2+ \sin^2\xi\cos^2\psi d\alpha^2\right.\nonumber \\
& &\left.-2\cos\xi\cos\psi\sin\xi\sin\psi d\xi d\psi \right)+e^{2\phi_2}\left(\cos^2\xi\sin^2\psi d\xi^2\right.\nonumber \\
& &+\cos^2\psi\sin^2\xi d\psi^2  +\sin^2\xi\sin^2\psi d\beta^2\nonumber \\
& &\left. \left.+2\cos\xi \cos\psi \sin\xi \sin\psi d\xi d\psi \right)\right]
\end{eqnarray}
where
\begin{equation}
\Delta=e^{4(\phi_1+\phi_2)}\cos^2\xi+e^{-2\phi_1}\sin^2\xi \cos^2\psi+e^{-2\phi_2}\sin^2\xi\sin^2\psi\, .
\end{equation}
The four-form field strength is much more complicated than the previous cases. We refrain from giving its explicit form here. The unbroken symmetry $SO(2)\times SO(2)$ corresponds to the isometry of $S^1\times S^1$ parametrized by coordinates $\alpha$ and $\beta$.
\\
\indent As $r\rightarrow \infty$, the solution becomes
\begin{equation}
\phi_1\sim\phi_2\sim e^{-\frac{4r}{L}}
\end{equation}
which again implies that $\phi_1$ and $\phi_2$ are dual to operators of dimension $\Delta=4$ in the dual $N=(2,0)$ SCFT. There are two possibilities for the IR behaviors.
\\
\indent As $g\rho\rightarrow 2C_1$, we have
\begin{eqnarray}
\phi_1&\sim&\phi_2\sim-\frac{1}{10}\ln(g\rho-2C_1),\\
A&=&-\frac{1}{2}\phi_1\sim\frac{1}{20}\ln(g\rho-2C_1).
\end{eqnarray}
Near the singularity, the scalar potential is unbounded above $V\rightarrow \infty$. The eleven-dimensional metric gives
\begin{equation}
\hat{g}_{00}\sim e^{\frac{5}{3}\phi_1}\rightarrow \infty\, .
\end{equation}
This singularity is then unphysical.
\\
\indent As $g\rho\rightarrow 2C_2$, we have
\begin{eqnarray}
\phi_1&\sim&-\frac{1}{5}\ln(g\rho-2C_2),\\
\phi_2&\sim&-\frac{3}{2}\phi_1\sim \frac{3}{10}\ln(g\rho-2C_2),\\
A&\sim&-\frac{1}{2}\phi_1\sim\frac{1}{10}\ln(g\rho-2C_2).
\end{eqnarray}
Near the singularity, we find $V\rightarrow -\infty$ and
\begin{equation}
\hat{g}_{00}\sim \textrm{constant}\, .
\end{equation}
In this case, the singularity is physical, and the solution describes an RG flow from $N=(2,0)$ SCFT to a non-conformal field theory in the IR with $SO(2)\times SO(2)$ symmetry.

\subsubsection{RG flow preserving $SO(3)$ symmetry}
In this case, the solution is more complicated. We will consider only a truncation of the full solution here. Making a consistent truncation by setting $\phi_3=0$, we obtain a simple solution to the truncated BPS equations
\begin{eqnarray}
A&=&\frac{1}{5}\phi_2-\frac{1}{4}\ln(1-e^{4\phi_2})+\frac{3}{20}\ln \left[1+C_1(e^{4\phi_2}-1)\right],\\
\phi_1&=&-\frac{1}{5}\phi_2+\frac{1}{10}\ln \left[1+C_1(e^{4\phi_2}-1)\right],\\
\phi_2&=&\frac{1}{2}\ln\left[\frac{1-e^{\frac{1}{2}(C-g\rho)}}{1+e^{\frac{1}{2}(C-g\rho)}}\right]
\end{eqnarray}
with $\frac{d\rho}{dr}=e^{6\phi_1}$.
\\
\indent Near the $AdS_7$ critical point in the UV as $r\rightarrow \infty$, we find, as in the previous cases,
\begin{equation}
\phi_1\sim\phi_2\sim e^{-\frac{4r}{L}},
\end{equation}
and, near the IR singularity as $g\rho\rightarrow C$, the solution becomes
\begin{eqnarray}
\phi_2&\sim&\frac{1}{2}\ln (g\rho-C),\\
\phi_1&\sim&-\frac{1}{5}\phi_2\sim-\frac{1}{10}\ln(g\rho-C),\\
A&\sim& \frac{1}{5}\phi_2\sim \frac{1}{10}\ln(g\rho-C).
\end{eqnarray}
In this case, the scalar potential diverges near the singularity $V\rightarrow \infty$, and the component of the eleven-dimensional metric gives
\begin{equation}
\hat{g}_{00}\sim e^{-\frac{2}{3}\phi_2}\rightarrow \infty\, .
\end{equation}
The singularity is then unphysical, and we will not give the corresponding eleven-dimensional solution in this case. It can be verified that a truncation with $\phi_2=0$ also gives similar result.
\\
\indent It should also be noted that in all of the above RG flows, there are only deformations by vacuum expectation values of the operators in agreement with the field theory results on the absence of deformations by turning on scalar operators corresponding, in the present case, to a non-normalizable mode $e^{-\frac{2r}{L}}$, see \cite{Intriligator_CFT_deform} for example. 

\subsection{Uplifted solutions to type IIA supergravity}
We now consider the uplift of the domain wall solutions in $CSO(4,0,1)$ gauge group to type IIA theory \cite{N4_7D_Henning}. Relevant parts of the truncation ansatz are reviewed in the appendix.
\\
\indent We first decompose the $SL(5)/SO(5)$ coset in term of the $SL(4)/SO(4)$ submanifold via
\begin{equation}\label{t=4decompose}
\mathcal{V}=e^{b_it^i}\widetilde{\mc{V}}e^{\phi_0t_0},
\end{equation}
where $\widetilde{\mc{V}}$ is the coset representative of $SL(4)/SO(4)\subset SL(5)/SO(5)$ coset. $t_0$ and $t^i$ correspond to the $SO(1,1)$ and four nilpotent generators in the decomposition $SL(5) \rightarrow SL(4) \times SO(1,1)$, respectively. With the coset representative \eqref{t=4decompose}, the scalar matrix $\mc{M}_{MN}$ takes the form of
\begin{equation}
\mc{M}_{MN}=\begin{pmatrix}
e^{-2\phi_0}\widetilde{\mc{M}}_{ij}+e^{8\phi_0}b_ib_j& e^{8\phi_0}b_i\\
e^{8\phi_0}b_j& e^{8\phi_0}
\end{pmatrix}
\label{SL5_SL4_coset}
\end{equation}
with $\widetilde{\mc{M}}=\widetilde{\mc{V}}\widetilde{\mc{V}}^T$. Relations between seven-dimensional fields and ten-dimensional ones are given in the appendix.
\\
\indent In all of the solutions considered here, we have $b_i=\chi_i=0$, so only the ten-dimensional metric, dilaton and three-form field strength are non-vanishing. The resulting solutions then, as expected for domain walls in seven dimensions, describe NS5-branes in the transverse space with different symmetries.

\subsubsection{Solution with $SO(4)$ symmetry}
In this case, we simply have $\widetilde{\mc{M}}_{ij}=\delta_{ij}$ and
\begin{eqnarray}
d\hat{s}^2_{10}&=&e^{\frac{3}{2}\phi_0}\left(e^{2A}dx^2_{1,5}+dr^2\right)+\frac{16}{g^2}e^{-\frac{5}{2}\phi_0}d\Omega^2_{(3)},\nonumber \\
\hat{F}_{(3)}&=&\frac{128}{g^3}\epsilon_{(3)},\qquad \hat{\varphi}=5\phi_0\, .
\end{eqnarray}
The solutions for $\phi_0$ and $A$ are given by
\begin{equation}
\phi_0=\frac{1}{2}\ln\left[\frac{gr}{10}+C\right]\qquad \textrm{and}\qquad A=\ln\left[\frac{gr}{10}+C\right].
\end{equation}
These are obtained from solving the BPS equations in \eqref{YSO(4)BPS1} and \eqref{YSO(4)BPS2} by renaming $\phi$ to $\phi_0$ and setting $\kappa=0$. We identify the resulting ten-dimensional solution with the ``near horizon'' geometry of NS5-branes in the transverse space $\mathbb{R}^4$.

\subsubsection{Solution with $SO(3)$ symmetry}
In this case, we parametrize the $SL(4)/SO(4)$ coset using
\begin{equation}
\widetilde{\mc{M}}_{ij}=\textrm{diag}(e^{2\phi},e^{2\phi},e^{2\phi},e^{-6\phi}).
\end{equation}
Solutions for scalars $\phi_0$ and $\phi$ can be obtained from the $SO(3)$ symmetric domain wall given in section \ref{SO3_DW_Y} by setting $\phi_2=0$ and using the relations
\begin{equation}
\phi_0=-\frac{1}{4}(3\phi_1+\phi_3)\qquad \textrm{and}\qquad \phi=\frac{1}{4}(5\phi_1-\phi_3).
\end{equation}
With $\kappa=0$ and $\sigma=1$, the domain wall solution is given by
\begin{eqnarray}
A&=&\frac{1}{5}\phi_3+\frac{3}{20}\ln(C_1+e^{4\phi_3}),\\
\phi_1&=&-\frac{1}{5}\phi_3+\frac{1}{10}\ln(C_1+e^{4\phi_3}),\\
2gC^{\frac{3}{5}}r&=&5e^{\frac{16}{5}\phi_3}\phantom{x}_2F_1\left(\frac{3}{5},\frac{4}{5},\frac{9}{5},-\frac{e^{4\phi_3}}{C_1}\right).
\end{eqnarray}
In this solution, $_2F_1$ is the hypergeometric function.
\\
\indent We now choose a specific form of the $S^3$ coordinates
\begin{equation}
\mu^i=(\sin\xi\hat{\mu}^a,\cos\xi),\qquad a=1,2,3
\end{equation}
with $\hat{\mu}^a$ being the coordinates on $S^2$ subject to the condition $\hat{\mu}^a\hat{\mu}^a=1$. With all these, we find the ten-dimensional fields
\begin{eqnarray}
d\hat{s}^2_{10}&=&e^{\frac{3}{2}\phi_0}\Delta^{\frac{1}{4}}\left(e^{2A}dx^2_{1,5}+dr^2\right)+\frac{16}{g^2}e^{-\frac{5}{2}\phi_0}\Delta^{-\frac{3}{4}}\left[\left(e^{-6\phi}\sin^2\xi+e^{2\phi}\cos^2\xi\right)d\xi^2\right. \nonumber \\
& &\left. +\sin^2\xi e^{2\phi}d\hat{\mu}^ad\hat{\mu}^a\right],\qquad e^{2\hat{\varphi}}=\Delta^{-1}e^{10\phi_0},\\
\hat{F}_{(3)}&=&\frac{64}{g^3}\Delta^{-2}\sin^3\xi\left(U\sin\xi d\xi+8e^{4\phi}\cos\xi \phi'dr\right)\wedge \epsilon_{(2)}
\end{eqnarray}
in which
\begin{eqnarray}
\Delta&=&e^{6\phi}\cos^2\xi+e^{-2\phi}\sin^2\xi,\qquad \epsilon_{(2)}=\frac{1}{2}\epsilon_{abc}\hat{\mu}^ad\hat{\mu}^b\wedge d\hat{\mu}^c,\nonumber \\
U&=&e^{12\phi}\cos^2\xi-e^{-4\phi}\sin^2\xi-e^{4\phi}(\sin^2\xi+3\cos^2\xi).
\end{eqnarray}
The unbroken $SO(3)$ symmetry corresponds to the isometry of $S^2\subset S^3$.

\subsubsection{Solution with $SO(2)\times SO(2)$ symmetry}
For $SO(2)\times SO(2)$ symmetric solutions, we use the following parametrization of $SL(4)/SO(4)$ coset
\begin{equation}
\widetilde{\mc{M}}_{ij}=\textrm{diag}(e^{2\phi},e^{2\phi},e^{-2\phi},e^{-2\phi}).
\end{equation}
In this case, the solutions for $\phi_0$ and $\phi$ can be obtained from the BPS equations given in section \ref{SO2_SO2_DW_Y} by setting $\sigma=1$, $\kappa=0$ and using the relations
\begin{equation}
\phi_0=-\frac{1}{2}(\phi_1+\phi_2)\qquad \textrm{and}\qquad \phi=\frac{1}{2}(\phi_1-\phi_2).
\end{equation}
The resulting seven-dimensional domain wall is given by
\begin{eqnarray}
A&=&\frac{1}{20}g\rho+\frac{1}{10}\ln(C_1+e^{\frac{1}{2}g\rho}),\\
\phi_1&=&C_2-\frac{1}{10}g\rho+\frac{3}{10}\ln(C_1+e^{\frac{1}{2}g\rho}),\\
\phi_2&=&C_2+\frac{3}{20}g\rho-\frac{1}{5}\ln(C_1+e^{\frac{1}{2}g\rho})
\end{eqnarray}
with $\frac{d\rho}{dr}=e^{-2\phi_2}$.
\\
\indent Choosing the coordinates on $S^3$ to be
\begin{equation}
\mu^i=(\cos\xi\cos\alpha,\cos\xi\sin\alpha,\sin\xi \cos\beta,\sin\xi\sin\beta),
\end{equation}
we find
\begin{eqnarray}
d\hat{s}^2_{10}&=&\Delta^{\frac{1}{4}}e^{\frac{3}{2}\phi_0}\left(e^{2A}dx^2_{1,5}+dr^2\right)
+\frac{16}{g^2}\Delta^{-\frac{3}{4}}e^{-\frac{5}{2}\phi_0}\left[(e^{2\phi}\sin^2\xi+e^{-2\phi}\cos^2\xi)d\xi^2
\right.\nonumber \\
& &\left.+e^{2\phi}\cos^2\xi d\alpha^2+e^{-2\phi}\sin^2\xi d\beta^2\right],\\
e^{2\hat{\varphi}}&=&\Delta^{-1}e^{10\phi_0},\qquad \hat{F}_{(3)}=\frac{128}{g^3}\Delta^{-2}\cos\xi \sin\xi d\alpha\wedge d\xi\wedge d\beta
\end{eqnarray}
with
\begin{equation}
\Delta=e^{-2\phi}\cos^2\xi+e^{2\phi}\sin^2\xi\, .
\end{equation}
In this case, the $SO(2)\times SO(2)$ symmetry corresponds to the isometry of $S^1\times S^1$ parametrized by coordinates $\alpha$ and $\beta$.

%%%%%%%%%%%%%%%%%%%%%%%%%%%%%%%%%%%%%%%%%%%%%%%%%%%%%%%%%%%%%%%%%%%%%%%%%%%%%%%%%%%%%%%%%%%%%%%%%%%%%%%%%%%%%%%%%%%%%%%%%%%%%%%%%%%%%%%%%
\section{Supersymmetric domain walls from gaugings in $\overline{\mathbf{40}}$ representation}\label{Z_gauging}
In this section, we consider gaugings with the embedding tensor in $\overline{\mathbf{40}}$ representation \cite{N4_7D_Henning}. Setting $Y_{MN}=0$, the quadratic constraint reads
\begin{equation}\label{RedQuadCon}
\epsilon_{MRSTU}Z^{RS,N}Z^{TU,P}=0.
\end{equation}
This condition can be solved by the following tensor
\begin{equation}\label{pureZ}
Z^{MN,P}=v^{[M}w^{N]P}
\end{equation}
with $w^{MN}=w^{(MN)}$. The $SL(5)$ symmetry can be used to fix $v^M=\delta^M_5$. It is useful to split the index $M=(i,5)$.
\\
\indent If, in addition, we set $w^{55}=w^{i5}=0$, the remaining $SL(4)$ symmetry, under which the vector $v^M=\delta^M_5$ is invariant, can be used to diagonalize $w^{ij}$. Accordingly, $w^{ij}$ can be written as
\begin{equation}\label{wij}
w^{ij}=\textrm{diag}(\underbrace{1,..,1}_p,\underbrace{-1,..,-1}_q,\underbrace{0,..,0}_r).
\end{equation}
The resulting gauge generators take the form of
\begin{equation}\label{pureZgaugeGen}
{(X_{ij})_k}^l=2\epsilon_{ijkm}w^{ml}
\end{equation}
which gives rise to $CSO(p,q,r)$ gauge group with $p+q+r=4$.
\\
\indent In these gaugings, following \cite{N4_7D_Henning}, it is convenient to parametrize the $SL(5)/SO(5)$ coset representative in term of $SL(4)/SO(4)$ submanifold as given in \eqref{SL5_SL4_coset}. After setting $Y_{MN}=0$ and using the inverse matrix $\mc{M}^{MN}$ of the form,  
\begin{equation}
\mc{M}^{MN}=\begin{pmatrix}e^{2\phi_0}\widetilde{\mc{M}}^{ij}& -e^{2\phi_0}b^i \\
-e^{2\phi_0}b^j& e^{-8\phi_0}+e^{2\phi_0}b_kb^k 
\end{pmatrix}
\end{equation}
with $\widetilde{\mc{M}}^{ij}$ being the inverse of $\widetilde{\mc{M}}_{ij}$ and $b^i=\widetilde{\mc{M}}^{ij}b_j$, we can rewrite the scalar Lagrangian as
\begin{equation}
e^{-1}\mc{L}_{\textrm{scalar}}=-8\pd_\mu \phi_0\pd^\mu\phi_0+\frac{1}{8}\pd_\mu \widetilde{\mc{M}}_{ij}\pd^\mu \widetilde{\mc{M}}^{ij}-\frac{1}{4}e^{10\phi_0}\widetilde{\mc{M}}^{ij}\pd_\mu b_i\pd^\mu b_j-V
\end{equation}
in which the scalar potential is given by
\begin{equation}\label{ZwithNilscalarPot}
V=\frac{g^2}{4}e^{14\phi_0}b_iw^{ij}\widetilde{\mc{M}}_{jk}w^{kl}b_l+\frac{g^2}{4}e^{4\phi_0}\left[2\widetilde{\mc{M}}_{ij}w^{jk}\widetilde{\mc{M}}_{kl}w^{li}-(\widetilde{\mc{M}}_{ij}w^{ij})^2\right].
\end{equation}
It should be noted that the nilpotent scalars $b_i$ appear quadratically in the Lagrangian, so setting them to zero is a manifestly consistent truncation.

\subsection{$SO(4)$ symmetric domain walls}
We firstly consider domain walls with the largest possible unbroken symmetry, $SO(4)\subset CSO(p,q,4-p-q)$. The only gauge group containing $SO(4)$ as a subgroup is $SO(4)$.
\\
\indent The embedding tensor is simply $w^{ij}=\delta^{ij}$, and there are no $SO(4)$ singlet scalars from $SL(4)/SO(4)$. We then take the coset representative to be $\widetilde{\mc{V}}=\mathbf{I}_4$. The scalar potential takes a particularly simple form
\begin{equation}\label{ZSO(4)Pot}
V=-2g^2e^{4\phi_0}\, .
\end{equation}
The Killing spinor still takes the fom \eqref{DW_Killing_spinor}, but unlike the previous cases, the appropriate projector for this type of gaugings is given by
\begin{equation}\label{pureZProj}
{(\Gamma_5)^a}_b\epsilon^b_0=-\gamma_r\epsilon^a_0\, .
\end{equation}
The appearance of $\Gamma^5$ rather than other $\Gamma^A$ with $A=1,2,3,4$ is due to the specific form of $v^M=\delta^M_5$ in the embedding tensor $Z^{MN,P}$.
\\
\indent It is now straightforward to derive the corresponding BPS equations
\begin{eqnarray}\label{ZSO(4)BPS}
A'&=& \frac{2g}{5}e^{-2\phi_0},\\
\phi_0'&=& \frac{g}{5}e^{-2\phi_0}.
\end{eqnarray}
We can readily find the solution
\begin{eqnarray}
\phi_0&=&\frac{1}{2}\ln\left[\frac{2gr}{5}+C\right],\\
A&=&\ln\left[\frac{2gr}{5}+C\right].
\end{eqnarray}

\subsection{$SO(3)$ symmetric domain walls}
We now look for more complicated solutions with $SO(3)$ symmetry. Gauge groups with an $SO(3)$ subgroup are $SO(4)$, $SO(3,1)$ and $CSO(3,0,1)$. We descibe them all at once by taking the symmetric matrix $w^{ij}$ in the form
\begin{equation}
w^{ij}=\textrm{diag}(1,1,1,\kappa)
\end{equation}
with $\kappa=1,-1,0$, respectively.
\\
\indent For simplicity, we truncate scalars $b_i$ out and consider only $\phi_0$ and $SL(4)/SO(4)$ scalars. With an explicit form of the $SL(4)/SO(4)$ coset representative
\begin{equation}
\widetilde{\mc{V}}=\textrm{diag}(e^{\phi},e^{\phi},e^{\phi},e^{-3\phi}),
\end{equation}
we obtain the scalar potential
\begin{equation}\label{ZSO(3)Pot}
V=-\frac{g^2}{4}e^{-4(\phi_0+3\phi)}(3e^{16\phi}+6\kappa e^{8\phi}+\kappa^2).
\end{equation}
Using the projector in \eqref{pureZProj}, we can derive the following set of BPS equations
\begin{eqnarray}\label{ZSO(3)BPS}
A'&=& \frac{g}{10}e^{-2(\phi_0+3\phi)}(3e^{8\phi}+\kappa),\\
\phi_0'&=& \frac{g}{20}e^{-2(\phi_0+3\phi)}(3e^{8\phi}+\kappa),\\
\phi'&=& -\frac{g}{4}e^{-2(\phi_0+3\phi)}(e^{8\phi}-\kappa).
\end{eqnarray}
The solutions for $A$ and $\phi_0$ are given by
\begin{eqnarray}
A&=&\frac{2}{5}\phi-\frac{1}{5}\ln(e^{8\phi}-\kappa),\\
\phi_0&=&\frac{1}{5}\phi-\frac{1}{10}\ln(e^{8\phi}-\kappa)+C_0\, .
\end{eqnarray}
The solution for $\phi(r)$ is given by
\begin{equation}
\phi=-\frac{5}{16}\ln\left[\frac{2}{5}(e^{-2C_0}gr-C)\right]
\end{equation}
for $\kappa=0$ and
\begin{equation}
4gr\kappa (e^{8\phi}-\kappa)^{\frac{1}{5}}=5e^{2C+\frac{32}{5}\phi}\left[4-3\left(1-\kappa e^{8\phi}\right)^{\frac{1}{5}}\phantom{x}_2F_1\left(\frac{1}{5},\frac{4}{5},\frac{9}{5},\kappa e^{8\phi}\right)\right]
\end{equation}
for $\kappa=\pm 1$. 

\subsection{$SO(2)\times SO(2)$ symmetric domain walls}
Possible domain wall solutions with $SO(2)\times SO(2)$ symmetry can be obtained from $SO(4)$ and $SO(2,2)$ gauge groups. These gauge groups are described by the component of the embedding tensor in the form of
\begin{equation}
w^{ij}=\textrm{diag}(1,1,\sigma,\sigma), \qquad \sigma=\pm 1\, .
\end{equation}
With the parametrization for the $SL(4)/SO(4)$ coset representative
\begin{equation}
\widetilde{\mc{V}}=\textrm{diag}(e^{\phi},e^\phi,e^{-\phi},e^{-\phi}),
\end{equation}
the scalar potential and the BPS equations are given by
\begin{equation}
V=-2g^2\sigma e^{-4\phi_0}
\end{equation}
and
\begin{eqnarray}
A'&=&\frac{1}{5}ge^{-2\phi_0-2\phi}(e^{4\phi}+\sigma),\\
\phi_0'&=&\frac{1}{10}ge^{-2\phi_0-2\phi}(e^{4\phi}+\sigma),\\
\phi'&=&\frac{1}{2}ge^{-2\phi_0-2\phi}(e^{4\phi}-\sigma).
\end{eqnarray}
The domain wall solution can be straightforwardly obtained
\begin{eqnarray}
A&=&2\phi_0,\\
\phi_0&=&\frac{1}{5}\phi-\frac{1}{10}\ln(e^{4\phi}-\sigma)+C_0,\\
6gr\sigma(e^{4\phi}-\sigma)^{\frac{1}{5}}&=&5e^{2C_0+\frac{12}{5}\phi}\left[3-2\left(1-\sigma e^{4\phi}\right)^{\frac{1}{5}}\phantom{x}_2F_1\left(\frac{1}{5},\frac{3}{5},\frac{8}{5},\sigma e^{4\phi}\right)\right].\qquad
\end{eqnarray}

\subsection{$SO(2)$ symmetric domain walls}
As a final example for domain wall solutions from gaugings in $\overline{\mathbf{40}}$ representation, we consider $SO(2)$ symmetric solutions. We again truncate out scalar fields $b_i$ and parametrize the $SL(4)/SO(4)$ coset representative as
\begin{equation}
\widetilde{\mc{V}}=e^{\phi_1Y_1+\phi_2Y_2+\phi_3Y_3}
\end{equation}
in which $Y_i$, $i=1,2,3$ are non-compact generators commuting with the $SO(2)$ symmetry generated by $X_{12}$. The explicit form of these generators is given by
 \begin{eqnarray}
Y_1&=& e_{11}+e_{22}-e_{33}-e_{44},\\
Y_2&=&e_{34}+e_{43},\\
Y_3&=&e_{33}-e_{44}\, .
\end{eqnarray}
There are many gauge groups admitting an $SO(2)$ subgroup. They are uniformly characterized by the following component of the embedding tensor
\begin{equation}\label{ZSO(2)w}
w^{ij}=\text{diag}(1,1,\sigma,\kappa).
\end{equation}
The scalar potential is computed to be
\begin{eqnarray}\label{ZSO(2)Pot}
V&=&-\frac{g^2}{16}e^{-4(\phi_0+\phi_1+\phi_3)}\left[ 8e^{4\phi_1+2\phi_3}\left[\kappa-\sigma+(\kappa+\sigma)\cosh{2\phi_2}\right]\right.\nonumber \\
& &-\left[\kappa-\sigma+(\kappa+\sigma)\cosh{2\phi_2}\right]^2-8e^{4\phi_1+6\phi_3}\left[\kappa-\sigma
-(\kappa+\sigma)\cosh{2\phi_2}\right]\nonumber \\
& &+e^{4\phi_3}\left[\kappa^2+10\kappa\sigma+\sigma^2-(\kappa+\sigma)^2\cosh{4\phi_2}\right]\nonumber \\
& &\left.+e^{8\phi_3}\left[\kappa-\sigma-(\kappa+\sigma)\cosh{2\phi_2}\right]^2\right].
\end{eqnarray}
It should be noted that the scalar potential for $CSO(2,0,2)$ gauge group with $\sigma=\kappa=0$ vanish identically. This leads to a Minkowski vacuum.
\\
\indent In this case, the BPS equations are much more complicated than those obtiained in the previous cases
\begin{eqnarray}\label{ZSO(2)BPS}
A'&=& \frac{1}{10}ge^{-2(\phi_0+\phi_1)}\left[2e^{4\phi_1}-(\kappa-\sigma)\sinh{2\phi_3}+(\kappa+\sigma)\cosh{2\phi_3}\cosh{2\phi_2}\right],\qquad \,\,\, \\
\phi_0'&=& \frac{1}{20}ge^{-2(\phi_0+\phi_1)}\left[2e^{4\phi_1}-(\kappa-\sigma)\sinh{2\phi_3}+(\kappa+\sigma)\cosh{2\phi_2}\cosh{2\phi_3}\right],\\
\phi_1'&=& -\frac{1}{4}ge^{-2(\phi_0+\phi_1)}\left[2e^{4\phi_1}+(\kappa-\sigma)\sinh{2\phi_3}-(\kappa+\sigma)\cosh{2\phi_2}\cosh{2\phi_3}\right],\\
\phi_2'&=& -\frac{1}{2}ge^{-2(\phi_0+\phi_1)}(\kappa+\sigma)\sinh{2\phi_2}\text{\ sech\ }{2\phi_3},\\
\phi_3'&=& \frac{1}{2}ge^{-2(\phi_0+\phi_1)}\left[(\kappa-\sigma)\cosh{2\phi_3}-(\kappa+\sigma)\cosh{2\phi_2}\sinh{2\phi_3}\right].
\end{eqnarray}
We are not able to completely solve these equations for arbitrary values of the parameters $\kappa$ and $\sigma$. However, the solutions can be separately found for various values of $\kappa$ and $\sigma$.

\subsubsection{Domain walls from $CSO(2,0,2)$ gauge group}
The simplest case is $CSO(2,0,2)$ gauge group corresponding to $\sigma=\kappa=0$. In this case, $\phi_2'=\phi_3'=0$ and the remaining BPS equations simplify considerably
\begin{equation}
A'=\frac{1}{5}ge^{-2\phi_0+\phi_1},\qquad
\phi_0'=\frac{1}{10}ge^{-2\phi_0+\phi_1},\qquad \phi_1'=-\frac{1}{2}ge^{-2\phi_0+\phi_1}\, .
\end{equation}
Scalars $\phi_2$ and $\phi_3$ can be consistently truncated out, and the solution for the remaining fields can be readily found
\begin{equation}
A=-\frac{1}{5}\phi_1,\qquad \phi_0=-\frac{1}{5}\phi_1+C_0,\qquad \phi_1=-\frac{5}{12}\ln\left[\frac{6}{5}(e^{-2C_0}gr-C)\right].
\end{equation}

\subsubsection{Domain walls from $SO(3,1)$ gauge group}
In this case, $\sigma=-\kappa=1$, and the BPS equations give $\phi_2'=0$. Similar to the previous case, $\phi_2$ does not appear in any BPS equations. After truncating out $\phi_2$, we find a domain wall solution
\begin{eqnarray}
\phi_1&=&\frac{1}{2}\phi_3-\frac{1}{4}\ln\left[1+C_1(1+e^{4\phi_3})\right],\\
\phi_0&=&C_0+\frac{1}{10}\phi_3-\frac{1}{10}\ln (1+e^{4\phi_3})+\frac{1}{20}\ln\left[1+C_1(1+e^{4\phi_3})\right],\\
\phi_3&=&\frac{1}{2}\ln\tan(C_3-g\rho),\\
A&=&2\phi_0
\end{eqnarray}
with $\rho$ defined by $\frac{d\rho}{dr}=e^{-2\phi_0-2\phi_1}$.

\subsubsection{Domain walls from $CSO(3,0,1)$ and $CSO(2,1,1)$ gauge groups}
In this case, we set $\kappa=0$ and $\sigma=\pm 1$ corresponding $CSO(3,0,1)$ and $CSO(2,1,1)$ gauge groups, respectively. All scalar fields are now non-vanishing. The domain wall solution is given by
\begin{eqnarray}
A&=&2\phi_0,\\
\phi_0&=&\frac{1}{20}\ln\left[\frac{1}{4}g\rho\left(C_0-g^2\rho^2e^{4C_1}-4e^{4C_1+C_3}g^2\rho^2-4e^{4C_1+2C_3}g^2\rho^2\right)\right],\\
\phi_1&=&C_1-5\phi_0-\frac{1}{4}\ln(1-e^{4\phi_2})+\frac{1}{4}\ln(1+2e^{C_3}+e^{2C_3}-e^{2C_3+4\phi_2}),\\
\phi_2&=&\frac{1}{4}\ln\left[\frac{4(1+e^{C_3})^2+(1+2e^{C_3})^2g^2\rho^2}{4e^{2C_3}+(1+2e^{C_3})^2g^2\rho^2}\right],\\
\phi_3&=&\frac{1}{4}\ln\left[\frac{(e^{2\phi_2}-1)(1+e^{C_3}+e^{C_3+2\phi_2})}{1+e^{C_3}+e^{2\phi_2}-e^{C_3+4\phi_2}}\right]
\end{eqnarray}
with $\frac{d\rho}{dr}=e^{-2\phi_0-2\phi_1}$. In this solution, we have shifted the coordinate $\rho$ to $\rho+\frac{C}{g \sigma}$ with $C$ being an integration constant in $\phi_2$ solution.

\subsubsection{Domain walls from $SO(4)$ and $SO(2,2)$ gauge groups}
In this case, we set $\kappa=\sigma=\pm 1$ corresponding to $SO(4)$ and $SO(2,2)$ gauge groups. The domain wall solution can be found as in the previous case
\begin{eqnarray}
A&=&2\phi_0,\\
\phi_0&=&C_0+\frac{1}{10}\ln\left[1+4e^{2C_3}-e^{-2g\sigma\rho}\right]+\frac{1}{40}(4\sigma+e^{20C_1}+4e^{20C_1+2C3})g\rho \nonumber \\
& &-\frac{\sigma}{160}e^{20C_1-2g\sigma\rho}(16e^{4(C_3+g\sigma\rho)}+8e^{2C_3+4g\sigma\rho}+e^{4g\sigma\rho}-1),\\
\phi_1&=&5C_1-\frac{1}{2}\ln(1-e^{4\phi_2})+\frac{1}{4}\ln\left[e^{2C_3}-e^{4\phi_2}-2e^{2C_3+4\phi_2}+e^{2C_3+8\phi_2}\right],\quad \\
\phi_2&=&\frac{1}{4}\ln\left[\frac{1-2e^{g\sigma \rho}+e^{2g\sigma \rho}+4e^{2C_3+2g\sigma \rho}}{1+2e^{g\sigma \rho}+e^{2g\sigma \rho}+4e^{2C_3+2g\sigma \rho}}\right],\\
\phi_3&=&\frac{1}{4}\ln\left[\frac{e^{2\phi_2}+e^{C_3+4\phi_2}-e^{C_3}}{e^{2\phi_2}+e^{C_3}-e^{C_3+4\phi_2}}\right]
\end{eqnarray}
with $\frac{d\rho}{dr}=e^{-2\phi_0-2\phi_1}$.

%%%%%%%%%%%%%%%%%%%%%%%%%%%%%%%%%%%%%%%%%%%%%%%%%%%%%%%%%%%%%%%%%%%%%%%%%%%%%%%%%%%%%%%%%%%%%%%%%%%%%%%%%%%%%%%%%%%%%%%%%%%%%%%%%%%%%%%%%
\section{Supersymmetric domain walls from gaugings in $\mathbf{15}$ and $\overline{\mathbf{40}}$ representations}\label{YZ_gauging}
We now consider gaugings with both components of the embedding tensor in $\mathbf{15}$ and $\overline{\mathbf{40}}$ representations non-vanishing. Following \cite{N4_7D_Henning}, we will choose a particular basis such that nonvanishing components of the embedding tensor are given by
\begin{equation}\label{mixembedding}
Y_{xy}, \qquad Z^{x\alpha,\beta}=Z^{x(\alpha,\beta)}, \qquad Z^{\alpha\beta,\gamma}
\end{equation}
in which the ranges of indices are given by $x=1,...,t$ and $\alpha=t+1,...,5$ for $t\equiv\textrm{rank} Y_{MN}$. We will also choose $Y_{xy}$ in the form
\begin{equation}\label{Yxy}
Y_{xy}=\text{diag}(\underbrace{1,..,1}_p,\underbrace{-1,..,-1}_q)
\end{equation}
with $p+q=t$. Tensors $Y_{xy}$, $Z^{x\alpha,\beta}$ and $Z^{\alpha\beta,\gamma}$ need to satisfy the quadratic constraint which is explicitly given by
\begin{equation}\label{mixQuadCon}
Y_{xy}Z^{y\alpha,\beta}+2\epsilon_{xMNPQ}Z^{MN,\alpha}Z^{PQ,\beta}=0\, .
\end{equation}
We will look for domain wall solutions in $SO(2,1)\ltimes \mathbf{R}^4$ and $SO(2)\ltimes \mathbf{R}^{4}$ gauge groups. The corresponding embedding tensors for these gauge groups have already been given in \cite{N4_7D_Henning}. We also emphasize that in the case of gaugings in $\mathbf{15}$ and $\overline{\mathbf{40}}$ representations, domain walls are $\frac{1}{4}$-BPS, preserving only eight supercharges. All gaugings in this case can be obtained from Scherk-Schwarz reduction of the maximal gauged supergravity in eight dimensions.

\subsection{$\frac{1}{4}$-BPS domain wall from $SO(2,1)\ltimes \mathbf{R}^4$ gauge group}
We begin with the $t=3$ case in which $Y_{xy}$ can be chosen to be $\textrm{diag}(1,1,\pm 1)$ \cite{N4_7D_Henning}. The component $Z^{\alpha\beta,\gamma}$ of the embedding tensor is not constrained by the quadratic constraint. Accordingly, $Z^{\alpha\beta,\gamma}$ does not affect the form of the gauge algebra and can be parametrized by an arbitrary two-component vector $v^\alpha$ as $Z^{\alpha\beta,\gamma}=\epsilon^{\alpha\beta}v^\gamma$. For simplicity, we will set $v^\alpha=0$. On the other hand, the quadratic constraint imposes the following condition on $Z^{x\alpha,\beta}$
\begin{equation}\label{mixt=3QuadCon}
\epsilon_{xyz}Z^{y\alpha,\gamma}\epsilon_{\gamma\delta}Z^{z\delta,\beta}=\frac{1}{8}Y_{xu}Z^{u\alpha,\beta}
\end{equation}
which implies that the $2\times 2$ matrices ${(\Sigma^x)_\alpha}^\beta=-16\epsilon_{\alpha\gamma}Z^{x\gamma,\beta}$ satisfy the algebra
\begin{equation}
[\Sigma^x,\Sigma^y]=2\epsilon^{xyu}Y_{uz}\Sigma^z\, .
\end{equation}
In terms of $\Sigma^x$, $Z^{x\alpha,\beta}$ component of the embedding tensor takes the form
\begin{equation}
Z^{x\alpha,\beta}=-\frac{1}{16}\epsilon^{\alpha\gamma}{(\Sigma^x)_\gamma}^\beta\, .
\end{equation}
\indent As pointed out in \cite{N4_7D_Henning}, a real, nonvanishing solution for $Z^{x\alpha,\beta}$ is possible only for $Y_{xy}$ generating a non-compact $SO(2,1)$ group. In this case, we take $Y_{xy}=\textrm{diag}(1,1,-1)$ and choose the explicit form for $\Sigma^x$ in terms of Pauli matrices as follow
\begin{equation}
\Sigma^1=\sigma_1,\qquad \Sigma^2=\sigma_3,\qquad \Sigma^3=i\sigma_2\, .
\end{equation}
The corresponding gauge generators are given by
\begin{equation}
{X_M}^N=\begin{pmatrix} \lambda^z{(t^z)_x}^y & Q^{(4)\beta}_{x} \\ 0_{2\times 3} & \frac{1}{2}\lambda^z {(\Sigma^z)_\alpha}^\beta  \end{pmatrix}
\end{equation}
with $\lambda^z\in \mathbb{R}$. It should be noted that the $SO(2,1)$ subgroup is embedded diagonally. The nilpotent generators $Q^{(4)\alpha}_{x}$ transform as $\mathbf{4}$ under $SO(2,1)$ and are obtained from projecting the tensor product $\mathbf{3}\otimes \mathbf{2}=\mathbf{2}+\mathbf{4}$ to representation $\mathbf{4}$. The resulting gauge group is then given by $SO(2,1)\ltimes \mathbf{R}^4$.
\\
\indent We will consider solutions that are invariant under the maximal compact subgroup $SO(2)\subset SO(2,1)$. Among the fourteen scalars in $SL(5)/SO(5)$ coset, there are four singlets corresponding to the following non-compact generators
\begin{eqnarray}\label{YZSO(2)diagYs}
\mathbf{Y}_1&=&2e_{1,1}+2e_{2,2}+2e_{3,3}-3e_{4,4}-3e_{5,5},\nonumber \\
\mathbf{Y}_2&=&e_{1,1}+e_{2,2}-2e_{3,3},\nonumber \\
\mathbf{Y}_3&=&e_{1,4}+e_{2,5}+e_{4,1}+e_{5,2},\nonumber \\
\mathbf{Y}_4&=&e_{1,5}-e_{2,4}-e_{4,2}+e_{5,1}\, .
\end{eqnarray}
With the $SL(5)/SO(5)$ coset representative
\begin{equation}\label{YZSO(2)diagYs}
\mathcal{V}=e^{\phi_0\mathbf{Y}_1+\phi_1\mathbf{Y}_2+\phi_2\mathbf{Y}_3+\phi_3\mathbf{Y}_4},
\end{equation}
we obtain the scalar potential
\begin{equation}\label{YZt=3scalarPot}
V=\frac{g^2}{64}e^{-2(4\phi_0-\phi_1)}\left[6\cosh{2\phi_2}\cosh{2\phi_3}+e^{6\phi_1}\right]
\end{equation}
which does not admit any critical points.
\\
\indent Contrary to the previous cases, finding the BPS equations in this case requires two projection conditions on the Killing spinors. In more detail, $A_1$ and $A_2$ tensors consist of two parts, one from $Y_{MN}$ and the other from $Z^{MN,P}$. The latter comes with an extra $SO(5)$ gamma matrices $\Gamma^A$ while the former does not. To obtain a consistent set of BPS equations, we impose the following projectors
\begin{equation}\label{YZProj}
\gamma_r\epsilon^a_0=-{(\Gamma_3)^a}_b\epsilon^b_0=\epsilon^a_0
\end{equation}
which reduce the number of supersymmetry to $\frac{1}{4}$ of the original amount or eight supercharges.
\\
\indent Following the same procedure as in the previous cases, we obtain the BPS equations
\begin{eqnarray}\label{YZt=3BPS}
A'&=& \frac{g}{40}e^{-2(2\phi_0+\phi_1)}\left(3\cosh{2\phi_2}\cosh{2\phi_3}-e^{6\phi_1}\right),\\
\phi_0'&=& \frac{g}{240}e^{-2(\phi_0+\phi_1)}\left(15\textrm{sech}{2\phi_2}\textrm{sech}{2\phi_3}-3\cosh{2\phi_2}\cosh{2\phi_3}-4e^{6\phi_1}\right),\quad\\
\phi_1'&=& \frac{g}{48}e^{-2(\phi_0+\phi_1)}\left(3\textrm{sech}{2\phi_2}\textrm{sech}{2\phi_3}+3\cosh{2\phi_2}\cosh{2\phi_3}+4e^{6\phi_1}\right),\\
\phi_2'&=&-\frac{3g}{16}e^{-2(2\phi_0+\phi_1)}\sinh{2\phi_2}\textrm{sech}{2\phi_3},\\
\phi_3'&=& -\frac{3g}{16}e^{-2(2\phi_0+\phi_1)}\cosh{2\phi_2}\sinh{2\phi_3}\, .
\end{eqnarray}
Introducing a new radial coordinate $\rho$ via $\frac{d\rho}{dr}=e^{-4\phi_0-2\phi_1}$, we can find a domain wall solution to these equations
\begin{eqnarray}
\phi_0&=&C_0+\frac{2}{45}(e^{3\phi_1}-3)\ln(1-e^{4\phi_2})-\frac{1}{60}\ln(e^{2C_3}-e^{4\phi_2}+e^{8\phi_2+2C_3}-2e^{4\phi_2+2C_3})\nonumber \\
& &-\frac{2}{45}\ln\left(1+e^{4\phi_2}+2\sqrt{e^{4\phi_2}-e^{2C_3}-e^{8\phi_2+2C_3}+2e^{4\phi_2+2C_3}}\right)\nonumber \\
& &+\frac{1}{6}\ln(1+e^{4\phi_2}),\\
\phi_1&=&C_1-5\phi_0-\ln(1-e^{4\phi_2})+\ln(1+e^{4\phi_2}),\\
\phi_2&=&\frac{1}{4}\ln\left[\frac{1+4e^{2C_3}-2e^{\frac{3}{8}g\rho}+e^{\frac{3}{4}g\rho}}{1+4e^{2C_3}+2e^{\frac{3}{8}g\rho}+e^{\frac{3}{4}g\rho}}\right],\\
\phi_3&=&\frac{1}{4}\ln\left[\frac{e^{2\phi_2}-e^{C_3}+e^{4\phi_2+C_3}}{e^{2\phi_2}+e^{C_3}-e^{4\phi_2+C_3}}\right],\\
A&=&\frac{1}{15}(e^{6\phi_1}-3)\ln(1-e^{4\phi_2})+\frac{1}{10}\ln\left(e^{2C_3}-e^{4\phi_2}+e^{2C_3+8\phi_2}-2e^{2C_3+4\phi_2}\right)\nonumber \\
& &-\frac{1}{15}e^{6\phi_1}\ln\left(2\sqrt{e^{4\phi_2}-e^{2C_3}-e^{8\phi_2+2C_3}+2e^{2C_3+4\phi_2}+e^{4\phi_2}+1}\right).
\end{eqnarray}

\subsection{$\frac{1}{4}$-BPS domain wall from $SO(2)\ltimes \mathbf{R}^{4}$ gauge group}
In this case, we have $t=2$ and $Y_{xy}=\delta_{xy}$, $x,y=1,2$. The quadratic constraint allows only the component $Z^{\alpha\beta,\gamma}$ to be non-vanishing. This component can be parametrized by a $3\times 3$ traceless matrix ${Z_\alpha}^\beta$, with ${Z_\alpha}^\alpha=0$, as
\begin{equation}\label{t=2Z}
Z^{\alpha\beta,\gamma}=\frac{1}{8}\epsilon^{\alpha\beta\delta}{Z_\delta}^\gamma\, .
\end{equation}
The corresponding gauge generators read
\begin{equation}
{X_M}^N=\begin{pmatrix} \lambda {t_x}^y & {Q_x}^\beta \\ 0_{3\times 2} & \lambda {Z_\alpha}^\beta  \end{pmatrix}
\end{equation}
with $\lambda\in \mathbb{R}$. ${t_x}^y=i\sigma_2$ generates the compact $SO(2)$ subgroup, and ${Q_x}^\alpha\in \mathbb{R}$ in general generate six translations $\mathbf{R}^6$ resulting in $SO(2)\ltimes \mathbf{R}^6$ gauge group. As pointed out in \cite{N4_7D_Henning}, the number of independent translations is reduced if there exist non-trivial solutions for $Q$ satisfying
\begin{equation}
tQ-QZ=0\, .
\end{equation}
\indent We will consider the compact case with $\textrm{Tr}Z^2=-2$. In this case, the gauged supergravity admits a half-supersymmetric ($N=2$) Minkowski vacuum, and the gauge group is reduced to $SO(2)\ltimes  \mathbf{R}^4\sim CSO(2,0,2)$. The $A_1$ tensor, related to the gravitino mass matrix, is given by
\begin{equation}
A_1^{ab}=-\frac{1}{20}e^{-6\phi_0}\cosh(2\phi_2)\cosh(2\phi_3)(\delta^a_1\delta^b_3-\delta^a_3\delta^b_1)
\end{equation}
which has only two zero eigenvalues indicating the supersymmetry breaking $N=4\rightarrow N=2$.
\\
\indent For definiteness, we take an explicit form of ${Z_\alpha}^\beta$ to be
\begin{equation}\label{ourt=2Z}
{Z_\alpha}^\beta=\begin{pmatrix} 0 & 0 & 0\\ 0 & 0 & -1 \\ 0 & 1 & 0 \end{pmatrix}.
\end{equation}
There are four $SO(2)$ singlet scalars corresponding to the following $SL(5)$ non-compact generators
\begin{eqnarray}\label{t=2SO(2)diagYs}
\overline{Y}_1&=& 3e_{1,1}+3e_{2,2}-2e_{3,3}-2e_{4,4}-2e_{5,5},\nonumber \\
\overline{Y}_2&=& e_{4,4}+e_{5,5}-2e_{3,3},\nonumber \\
\overline{Y}_3&=& e_{1,4}+e_{2,5}+e_{4,1}+e_{5,2},\nonumber \\
\overline{Y}_4&=& e_{1,5}-e_{2,4}-e_{4,2}+e_{5,1}\, .
\end{eqnarray}
Using the parametrization of the $SL(5)/SO(5)$ coset representative in the form
\begin{equation}
\mathcal{V}=e^{\phi_0\overline{Y}_1+\phi_1\overline{Y}_2+\phi_2\overline{Y}_3+\phi_3\overline{Y}_4},
\end{equation}
we find that the scalar potential vanishes identically. This is in agreement with $CSO(2,0,2)$ gauge group considered in the previous section.
\\
\indent With the projector \eqref{YZProj}, we can derive the following BPS equations
\begin{eqnarray}\label{YZt=2BPS}
A'&=&\frac{g}{10}e^{-6\phi_0}\cosh{2\phi_2}\cosh{2\phi_3},\\
\phi_0'&=& \frac{g}{60}e^{-6\phi_0}\left(\cosh{2\phi_2}\cosh{2\phi_3}+5\textrm{sech}{2\phi_2}\textrm{sech}{2\phi_3}\right),\\
\phi_1'&=& \frac{g}{12}e^{-6\phi_0}\left(\cosh{2\phi_2}\cosh{2\phi_3}-\textrm{sech}{2\phi_2}\textrm{sech}{2\phi_3}\right),\\
\phi_2'&=& -\frac{g}{4}e^{-6\phi_0}\sinh{2\phi_2}\textrm{sech}{2\phi_3},\\
\phi_3'&=& -\frac{g}{4}e^{-6\phi_0}\cosh{2\phi_2}\sinh{2\phi_3}\, .
\end{eqnarray}
By using a new radial coordinate $\rho$ defined by $\frac{d\rho}{dr}=e^{-6\phi_0}$, we find a domain wall solution to the above equations
\begin{eqnarray}
\phi_0&=&C_0-\frac{1}{5}\ln(1-e^{4\phi_2})+\frac{1}{6}\ln(1+e^{4\phi_2})\nonumber \\
& &+\frac{1}{60}\ln\left[e^{2C_3}-e^{4\phi_2}+e^{2C_3+8\phi_2}-2e^{2C_3+4\phi_2}\right],\\
\phi_1&=&C_1-\frac{1}{6}\ln(1+e^{4\phi_2})+\frac{1}{12}\ln\left[e^{2C_3}-e^{4\phi_2}+e^{2C_3+8\phi_2}-2e^{2C_3+4\phi_2}\right],\quad\\
\phi_2&=&\frac{1}{4}\ln\left[\frac{1+4e^{2C_3}-2e^{\frac{1}{2}g\rho}+e^{g\rho}}{1+4e^{2C_3}+2e^{\frac{1}{2}g\rho}+e^{g\rho}}\right],\\
\phi_3&=&\frac{1}{4}\ln\left[\frac{e^{2\phi_2}+e^{4\phi_2+C_3}-e^{C_3}}{e^{2\phi_2}-e^{4\phi_2+C_3}+e^{C_3}}\right],\\
A&=&-\frac{1}{5}\ln(1-e^{4\phi_2})+\frac{1}{10}\ln\left[e^{2C_3}-e^{4\phi_2}+e^{2C_3+8\phi_2}-2e^{2C_3+4\phi_2}\right].
\end{eqnarray}

%%%%%%%%%%%%%%%%%%%%%%%%%%%%%%%%%%%%%%%%%%%%%%%%%%%%%%%%%%%%%%%%%%%%%%%%%%%%%%%%%%%%%%%%%%%%%%%%%%%%%%%%%%%%%%%%%%%%%%%%%%%%%%%%%%%%%%%%%
\section{Conclusions and discussions}\label{conclusion}
We have studied supersymmetric domain walls in $N=4$ gauged supergravity in seven dimensions with various gauge groups. There are both half-supersymmetric and $\frac{1}{4}$-supersymmetric solutions depending on which components of the embedding tensor in the $\mathbf{15}$ and $\overline{\mathbf{40}}$ representations of the global symmetry $SL(5)$ lead to the gauging.
\\
\indent For $SO(5)$ gauge group, the gauged supergravity admits a supersymmetric $AdS_7$ vacuum and can be embedded in eleven-dimensional supergravity. Accordingly, there exist domain walls that are asymptotic to the $AdS_7$ vacuum and can be interpreted as RG flows from $N=(2,0)$ SCFT, dual to the $AdS_7$, to non-conformal field theories in the IR. The resulting solutions can be uplifted to eleven dimensions. Furthermore, solutions from $CSO(4,0,1)$ gauged supergravity can be embedded in type IIA theory via a consistent $S^3$ truncation. These solutions with clear higher-dimensional origins would be useful in the study of the AdS/CFT correspondence and various dynamical aspects of M5-branes and NS5-branes in different transverse spaces.
\\
\indent There are a number of future directions to pursue. First of all, it is interesting to look for domain walls from $CSO(1,0,4)$ and $CSO(1,0,3)$ gauge groups that would presumably involve many non-vanishing scalars. These are called elementary domain walls in \cite{Eric_SUSY_DW}. With the truncation ansatz given in \cite{Henning_Emanuel}, it would be of particular interest to uplift the solutions from $SO(4)$ gauged supergravity to type IIB theory and study the field theory on the world-volume of NS5- and D5-branes. Using the solutions from $SO(5)$ and $CSO(4,0,1)$ gauged supergravities given here to holographically study field theories on M5-branes and NS5-branes also deserves further investigation along the line of \cite{KS,MN_1,6D_flow_Tomasiello}. Finally, finding supersymmetric domain walls with non-vanishing vector and tensor fields as in half-maximal gauged supergravity studied in \cite{7D_sol_Dibitetto,6D_surface_Dibitetto,7D_N2_DW_3_form} is worth considering.
\vspace{0.5cm}\\
%%%%%%%%%%%%%%%%%%%%%%%%%%%%%%%%%%%%%%%%%%%%%%%%%%%%%%%%%%%%%%%%%%%%%%%%%%%%%%%%%%%%%%%%%%%%%%%%%%%%%%%%%%%%%%%%%%%%%%%%%%%%%%%%%%%%%%%%%
{\large{\textbf{Acknowledgement}}} \\
We would like to thank Nikolay Bobev and Emanuel Malek for useful comments and correspondences. This work is supported by The Thailand Research Fund (TRF) under grant RSA6280022.
%%%%%%%%%%%%%%%%%%%%%%%%%%%%%%%%%%%%%%%%%%%%%%%%%%%%%%%%%%%%%%%%%%%%%%%%%%%%%%%%%%%%%%%%%%%%%%%%%%%%%%%%%%%%%%%%%%%%%%%%%%%%%%%%%%%%%%%%%%%%%%%%%%%%%%%%%%%%%%%%
\appendix
\section{Truncation ansatze}
In this appendix, we collect some useful formulae for truncations of eleven-dimensional supergravity on $S^4$ and type IIA theory on $S^3$. The former leads to $SO(5)$ gauged supergravity while the latter gives $CSO(4,0,1)$ gauged supergravity in seven dimensions. The complete $S^4$ truncation has been constructed in \cite{11D_to_7D_Nastase1,11D_to_7D_Nastase2}, but we will use the convention of \cite{S3_S4_typeIIA}. Apart from some notational changes, this appendix closely follows \cite{S3_S4_typeIIA} to which the reader is referred for more detail. Since the seven-dimensional solutions considered here do not involve vector and tensor fields, we will only give the truncation ansatze with only seven-dimensional metric and scalars non-vanishing for brevity.

\subsection{Eleven-dimensional supergravity on $S^4$}
The ansatz for the eleven-dimensional metric is given by
\begin{equation}
d\hat{s}^2_{11}=\Delta^{\frac{1}{3}}ds^2_7+\frac{1}{\hat{g}^2}\Delta^{-\frac{2}{3}}T^{-1}_{MN}d\mu^Md\mu^N
\end{equation}
with the coordinates $\mu^M$, $M=1,2,3,4,5$, on $S^4$ satsifying $\mu^M\mu^M=1$. $T_{MN}$ is a unimodular $5\times 5$ symmetric matrix describing scalar fields in the $SL(5)/SO(5)$ coset. The warped factor is given by
\begin{equation}
\Delta=T_{MN}\mu^M\mu^N\, .
\end{equation}
The ansatz for the four-form field strength reads
\begin{eqnarray}
\hat{F}_{(4)}&=&\frac{1}{\hat{g}^3}\Delta^{-2}\left[-U\epsilon_{(4)}+\frac{1}{3!}\epsilon_{M_1\ldots M_5}\mu^M\mu^NT^{M_1M}dT^{M_2N}\wedge d\mu^{M_3}\wedge d\mu^{M_4}\wedge d\mu^{M_5}\right]\nonumber \\
& &
\end{eqnarray}
with the following definitions
\begin{eqnarray}
U&=&2T_{MN}T_{NP}\mu^{M}\mu^P-\Delta T_{MM}, \\
\epsilon_{(4)}&=&\frac{1}{4!}\epsilon_{M_1\ldots M_5}\mu^{M_1}d\mu^{M_2}\wedge d\mu^{M_3}\wedge d\mu^{M_4} \wedge d\mu^{M_5}\, .
\end{eqnarray}
After multiplied by $\frac{1}{2}$, the seven-dimensional Lagrangian can be written as
\begin{equation}
e^{-1}\mc{L}_{S^4}=\frac{1}{2}R+\frac{1}{8}\pd_\mu T^{-1}_{MN}\pd^\mu T_{MN}-\frac{1}{4}\hat{g}^2\left[2T_{MN}T_{MN}-(T_{MM})^2\right].
\end{equation}
Comparing with \eqref{7D_Lgauge} and setting $Y_{MN}=\delta_{MN}$, $Z^{MN,P}=0$, we find the following identification
\begin{equation}
T_{MN}=\mc{M}^{MN}\qquad \textrm{and}\qquad \hat{g}=\frac{1}{4}g\, .\label{7D_higher_D_rel}
\end{equation}

\subsection{Type IIA supergravity on $S^3$}
By taking a limit in which the four-sphere $S^4$ degenerates to $\mathbb{R}\times S^3$ followed by a standard Kaluza-Klein reduction on $S^1$, a consistent truncation of type IIA supergravity on $S^3$ has been obtained in \cite{S3_S4_typeIIA}. To present this ansatz, we will split the index $M$ as $M=(i,5)$, $i=1,2,3,4$. The $SL(5)/SO(5)$ coset is decomposed under the $SL(4)/SO(4)$ submanifold as
\begin{equation}
T^{-1}_{MN}=\begin{pmatrix}  \Phi^{-\frac{1}{4}}M^{-1}_{ij}+\Phi \chi_i\chi_j& \Phi \chi_i \\
\Phi \chi_j &\Phi
\end{pmatrix}\label{Ti_S3}
\end{equation}
where $M_{ij}$ is a unimodular $4\times 4$ symmetric matrix describing the $SL(4)/SO(4)$ coset.
\\
\indent The ten-dimensional metric, dilaton and various form field strength tensors are given by
\begin{eqnarray}
d\hat{s}^2_{10}&=&\Phi^{\frac{3}{16}}\Delta^{\frac{1}{4}}ds^2_7+\frac{1}{\hat{g}^2}\Phi^{-\frac{5}{16}}\Delta^{-\frac{3}{4}}M^{-1}_{ij}d\mu^id\mu^j,\\
e^{2\hat{\varphi}}&=&\Delta^{-1}\Phi^{\frac{5}{4}},\qquad  \hat{F}_{(2)}= d\chi_i\wedge d\mu^i,\nonumber \\
\hat{F}_{(3)}&=&\frac{1}{\hat{g}^3}\Delta^{-2}\left[-U\epsilon_{(3)}+\frac{1}{2}\epsilon_{i_1i_2i_3i_4}M_{i_1j}\mu^j\mu^kdM_{i_2k}\wedge d\mu^{i_3}\wedge d\mu^{i_4}\right],\\
\hat{F}_{(4)}&=&\frac{1}{\hat{g}^3}\Delta^{-1}M_{ij}\mu^jd\chi_i\wedge \epsilon_{(3)}
\end{eqnarray}
with
\begin{eqnarray}
\epsilon_{(3)}&=&\frac{1}{3!}\epsilon_{ijkl}\mu^id\mu^j\wedge d\mu^k\wedge d\mu^l,\\
U&=&2M_{ij}M_{jk}\mu^i\mu^k-\Delta M_{ii}\, .
\end{eqnarray}
Using the relation \eqref{7D_higher_D_rel} and comparing the $SL(5)/SO(5)$ coset given in \eqref{SL5_SL4_coset} with \eqref{Ti_S3}, we find the relations
\begin{equation}
\Phi=e^{8\phi_0},\qquad \chi_i=b_i,\qquad M^{-1}_{ij}=\widetilde{\mc{M}}_{ij}\, .
\end{equation}
In this case, $\mu^i$ is the coordinates on $S^3$ satisfying $\mu^i\mu^i=1$. The gauge coupling $\hat{g}$ is related to $g$ by $\hat{g}=\frac{1}{4}g$ as in the $S^4$ truncation of eleven-dimensional supergravity.

%%%%%%%%%%%%%%%%%%%%%%%%%%%%%%%%%%%%%%%%%%%%%%%%%%%%%%%%%%%%%%%%%%%%%%%%%%%%%%%%%%%%%%%%%%%%%%%%%%%%%%%%%%%%%%%%%%%%%%%%%%%%%%%%%%%%%%%%%


\begin{thebibliography}{99}
\bibitem{DW_QFT1} H.J. Boonstra, K. Skenderis and P.K. Townsend, ``The domain-wall/QFT
correspondence'', JHEP 01 (1999) \textbf{003}, arXiv: hep-th/9807137.
\bibitem{DW_QFT2} T. Gherghetta and Y. Oz, ``Supergravity, Non-Conformal Field Theories and
Brane-Worlds'', Phys. Rev. \textbf{D65} (2002) 046001, arXiv: hep-th/0106255.
\bibitem{DW_QFT3} Ingmar Kanitscheider, Kostas Skenderis and Marika Taylor, ``Precision holography for
non-conformal branes'', JHEP 09 (2008) \textbf{094}, arXiv: 0807.3324.
\bibitem{maldacena} J. M. Maldacena, ``The large $N$ limit of
superconformal field theories and supergravity'', Adv. Theor. Math.
Phys. \textbf{2} (1998) 231-252, arXiv: hep-th/9711200.
\bibitem{DW_cosmology1} K. Skenderis and P. K. Townsend, ``Hidden supersymmetry of domain walls and
cosmologies'', Phys. Rev. Lett. \textbf{96} (2006) 191301, arXiv: hep-th/0602260.
\bibitem{DW_cosmology2} Kostas Skenderis and Paul K. Townsend, ``Hamilton-Jacobi method for Domain Walls
and Cosmologies'', Phys. Rev. \textbf{D74} (2006) 125008, arXiv: hep-th/0609056.
\bibitem{DW_cosmology3} Kostas Skenderis, Paul K. Townsend and Antoine Van Proeyen,
``Domain-wall/Cosmology correspondence in AdS/dS supergravity'', JHEP 08 (2007)
\textbf{036}, arXiv: 0704.3918.
\bibitem{counting_DW} A. Kleinschmidt, ``Counting supersymmetric branes'', JHEP 10
(2011) \textbf{144}, arXiv: 1109.2025.
\bibitem{E11_West} P. C. West, ``E(11) and M theory'', Class. Quant. Grav. \textbf{18} (2001) 4443,
arXiv: hep-th/0104081.
\bibitem{Eric_DW_10D} E. Bergshoeff, M. de Roo, M. B. Green, G. Papadopoulos and
P. K. Townsend, ``Duality of type II 7 branes and 8 branes'', Nucl. Phys. \textbf{B470} (1996) 113, arXiv: hep-th/9601150.
\bibitem{9D_DW_Cowdall} P. M. Cowdall, ``Novel Domain Wall and Minkowski Vacua of $D=9$ Maximal $SO(2)$ Gauged Supergravity'', Nucl. Phys. \textbf{B600} (2001) 81, arXiv: hep-th/0009016.
\bibitem{9D_DW} E. Bergshoeff, U. Gran and D. Roest, ``Type IIB seven-brane solutions
from nine-dimensional domain walls'', Class. Quant. Grav. \textbf{19} (2002) 4207, arXiv: hep-th/0203202.
\bibitem{8D_DW1} N. Alonso Alberca, E. Bergshoeff, U. Gran, R. Linares, T. Ortin and
D. Roest, ``Domain walls of $D = 8$ gauged supergravities and their $D =
11$ origin'', JHEP 06 (2003) \textbf{038}, arXiv: hep-th/0303113.
\bibitem{8D_DW2} E. Bergshoeff, U. Gran, R. Linares, M. Nielsen, T. Ortin and D. Roest,
``The Bianchi classification of maximal D = 8 gauged supergravities'', Class. Quant. Grav. \textbf{20} (2003) 3997, arXiv: hep-th/0306179.
\bibitem{Pope_DW_massive} P. M. Cowdall, H. Lu, C. N. Pope, K. S. Stelle and P. K. Townsend,
``Domain walls in massive supergravities'', Nucl. Phys. \textbf{B486} (1997) 49, arXiv: hep-th/9608173.
\bibitem{DW_algebraic_curve} I. Bakas, A. Brandhuber and K. Sfetsos, ``Domain walls of gauged supergravity, M-branes, and algebraic curves'', Adv. Theor. Math. Phys. \textbf{3} (1999) 1657-1719, arXiv: hep-th/9912132.
\bibitem{Eric_DW_maximal_SUGRA} E. Bergshoeff, M. Nielsen and D. Roest, ``The Domain Walls of Gauged
Maximal Supergravities and their M-theory Origin'', JHEP 07 (2004) \textbf{006}, arXiv: hep-th/0404100.
\bibitem{Eric_SUSY_DW} E. A. Bergshoeff, A. Kleinschmidt and F. Riccioni, ``Supersymmetric Domain
Walls'', Phys. Rev. \textbf{D86} (2012) 085043, arXiv: 1206.5697.
\bibitem{Pope_symmetric_poten} M Cvetic, S. S. Gubser, H. Lu and C. N. Pope, ``Symmetric Potentials of Gauged Supergravities in Diverse Dimensions and Coulomb Branch of Gauge Theories'', Phys. Rev. \textbf{D62} (2000) 086003, arXiv: hep-th/9909121. 
\bibitem{DW_Hull} C. M. Hull, ``Domain Wall and de Sitter Solutions of Gauged Supergravity'', JHEP 11 (2011) \textbf{061}, arXiv: hep-th/0110048.
\bibitem{4D_DW} H. Singh, ``New Supersymmetric Vacua for N=4, D=4 Gauged Supergravity'', Phys. Lett. \textbf{B429} (1998) 304-312, arXiv: hep-th/9801038.
\bibitem{3DN4_DW} P. Karndumri, ``Domain walls in three dimensional gauged supergravity'', JHEP 10 (2012) \textbf{001}, arXiv: 1207.1227.
\bibitem{3DN10_DW} P. Karndumri, ``$\frac{1}{2}$-BPS Domain wall from $N=10$ three dimensional gauged supergravity'', JHEP 11 (2013) \textbf{023}, arXiv: 1307.6641.
\bibitem{2D_DW1} T. Ortiz and H. Samtleben, ``$SO(9)$ supergravity in two dimensions'', JHEP 01 (2013) \textbf{183}, arXiv: 1210.4266.
\bibitem{2D_DW2} A. Anabalon, T. Ortiz and H. Samtleben, ``Rotating D0-branes and consistent truncations of supergravity'', Phys. Lett. \textbf{B727} (2013) 516-523, arXiv: 1310.1321.
\bibitem{7DN4_gauged1} M. Pernici, K. Pilch and P. van Nieuwenhuizen, ``Gauged maximally extended
supergravity in seven-dimensions'', Phys. Lett. \textbf{B143} (1984) 103.
\bibitem{7DN4_gauged2} M. Pernici, K. Pilch, P. van Nieuwenhuizen and N. P. Warner, ``Noncompact gaugings and critical points of maximal supergravity in seven-dimensions'', Nucl. Phys. \textbf{B249} (1985) 381.
\bibitem{11D_to_7D_Townsend} K. Pilch, P. van Nieuwenhuizen, and P. K. Townsend, ``Compactification of $d = 11$
Supergravity on $S(4)$ (Or $11 = 7 + 4$, Too)'', Nucl. Phys. \textbf{B242} (1984) 377–392.
\bibitem{11D_to_7D_Nastase1} H. Nastase, D. Vaman, and P. van Nieuwenhuizen, ``Consistent nonlinear KK reduction of
$11-d$ supergravity on $AdS(7) \times  S(4)$ and selfduality in odd dimensions'', Phys. Lett. \textbf{B469} (1999) 96–102, arXiv:hep-th/9905075.
\bibitem{11D_to_7D_Nastase2} H. Nastase, D. Vaman, and P. van Nieuwenhuizen, Consistency of the $AdS_7 \times S^4$ reduction and the origin of self-duality in odd dimensions, Nucl. Phys. \textbf{B581} (2000) 179–239, [hep-th/9911238].
\bibitem{N4_7D_Henning} H. Samtleben and M. Weidner, ``The maximal $D=7$ supergravities'', Nucl. Phys. \textbf{725} (2005) 383-419, arXiv: hep-th/0506237.
\bibitem{S3_S4_typeIIA} M. Cvetic, H. Lu, C. N. Pope, A. Sadrzadeh and T. A. Tran, ``$S^3$ and $S^4$ reductions of type IIA supergravity, Nucl. Phys. \textbf{B590} (2000) 233–251, arXiv: hep-th/0005137.
\bibitem{Henning_Emanuel} E. Malek and H. Samtleben, ``Dualising consistent IIA /IIB truncations'', JHEP 12 (2015) \textbf{029}, arXiv: 1510.03433.
\bibitem{warner} N. P. Warner, ``Some New Extrema of the Scalar Potential of Gauged $N = 8$ Supergravity'', Phys. Lett. \textbf{B128} (1983) 169.
\bibitem{non-CFT_flow_Zaffaroni} M. Petrini and A. Zaffaroni, ``The holographic RG flow to conformal and
non-conformal theory'', arXiv: hep-th/0002172.
\bibitem{Gubser_non_CFT} S. S. Gubser, ``Non-conformal examples of AdS/CFT'', Class. Quant. Grav. \textbf{17} (2000) 1081-1092, arXiv: hep-th/9910117.
\bibitem{Gubser_singularity} S. S. Gubser, ``Curvature singularities: the good, the bad and the naked'', Adv. Theor.
Math. Phys. \textbf{4} (2000) 679-745.
\bibitem{maldacena_nogo} J. Maldacena and C. Nunez, ``Supergravity description of field theories on
curved manifolds and a no go theorem'', Int. J. Mod. Phys. \textbf{A16} (2001) 822,
arXiv: hep-th/0007018.
\bibitem{Bobev_6D_holography} N. Bobev, G. Dibitetto, F. F. Gautason and B. Truijen, ``Holography, Brane Intersections and Six-dimensional SCFTs'', JHEP 02 (2017) \textbf{116}, arXiv: 1612.06324.
\bibitem{Intriligator_CFT_deform} C. Cordova, T. T. Dumitrescu and K. Intriligator, ``Deformations of Superconformal Theories'', JHEP 11 (2016) \textbf{135}, arXiv: 1602.01217.  
\bibitem{KS} I. R. Klebanov and M. J. Strassler, ``Supergravity and a Confining Gauge Theory: Duality Cascades and $\chi$SB-Resolution of Naked Singularities'', JHEP 08 (2000) \textbf{052}, arXiv: hep-th/0007191. 
\bibitem{MN_1} J. M. Maldacena and C. Nunez, ``Towards the large N limit of pure $N=1$ super Yang Mills'', Phys. Rev. Lett. \textbf{86} (2001) 588-591, arXiv: hep-th/0008001.
\bibitem{6D_flow_Tomasiello} G. B. De Luca, A. Gnecchi, G. Lo Monaco and A. Tomasiello, ``Holographic duals of 6d RG flows'', JHEP 03 (2019) \textbf{035}, arXiv: 1810.10013.
\bibitem{7D_sol_Dibitetto} G. Dibitetto and N. Petri, ``BPS objects in $D=7$ supergravity and their M-theory 
origin'', JHEP 12 (2017) \textbf{041}, arXiv: 1707.06152.
\bibitem{6D_surface_Dibitetto} G. Dibitetto and N. Petri, ``6d surface defects from massive type IIA'', JHEP 01 (2018) \textbf{039}, arXiv: 1707.06154.
\bibitem{7D_N2_DW_3_form} P. Karndumri and P. Nuchino, ``Supersymmetric solutions from matter-coupled $7D$ $N=2$ gauged supergravity'', Phys. Rev. \textbf{D98} (2018) 086012, arXiv: 1806.04064.
\end{thebibliography}
\end{document}